\documentclass[conference]{IEEEtran}
\usepackage[]{wspr}

\usepackage{amsmath}
\usepackage{tcolorbox}
\usepackage{algpseudocode,algorithm}
\usepackage{mdframed}
\usepackage{etoolbox}
\usepackage{xcolor}
\usepackage{mathtools}
\usepackage{adjustbox}
\usepackage{tikz}
\usepackage{amsthm}
\usepackage{enumitem}
\usepackage{setspace}
\usepackage{pgfplots}

\usetikzlibrary{positioning,calc,arrows.meta,decorations.markings}

\tikzset{
  actor/.style={
    draw=black, thick, rectangle,
    fill=white,
    inner sep=4pt,
    align=center, 
    anchor=north,
    minimum height=0.8cm,
    font=\footnotesize,
  },
  protobox/.style={
    draw=black, thick, rectangle, rounded corners,
    fill=white,
    inner sep=4pt,
    align=left, anchor=north
  },
  lifeline/.style={
    circle, 
    fill, 
    inner sep=2pt
  },
  msg req/.style={
    postaction={
      decorate,
      decoration={
        markings,
        mark=at position 0 with {\draw[solid,fill=white] circle[radius=2pt];}
      }
    }
  },
  msg res/.style={
    postaction={
      decorate,
      decoration={
        markings,
        mark=at position 1 with {
          \draw[draw,solid,fill=white] circle[radius=2pt];
        }
      }
    }
  },
}
\newcommand{\protoboxtitle}[1]{\textbf{\underline{{\footnotesize #1}}}\par}

\usepackage[
	lambda,
	operators,
	advantage,
	sets,
	adversary, 
	landau,
	probability,
	notions,
	logic,
	ff,
	mm,
	primitives,
	events,
	complexity,
	asymptotics,
	keys]{cryptocode}

\newcommand{\circletext}[2]{\tikz[baseline=(char.base)]{
    \node[shape=circle, draw=#1, fill=#1, text=white, inner sep=0.5pt] (char) {\footnotesize #2};
  }}

\newtheorem{theorem}{Theorem}

\newif\ifoobfullversion

\newcommand{\sysname}{Sidecar\xspace}
\newcommand{\stirsha}{S/S\xspace}
\newcommand{\oobss}{OOB-\stirsha}
\newcommand{\thislib}{\texttt{libsidecar}\xspace}

\newcounter{mysubsection}
\newcommand{\mysubsection}[1]{\subsection{#1}
}

\preto\section{\setcounter{mysubsection}{0}}
\preto\subsection{\setcounter{mysubsection}{0}}

\newcommand{\mysubsubsection}[1]{\vspace{4pt}\noindent\textbf{#1.}}
\newcommand{\myquestion}[1]{\vspace{0pt}\noindent\textbf{#1?}}

\definecolor{VerifyLightRed}{RGB}{255, 104, 103}

\newcommand{\fstirsha}{STIR/SHAKEN\xspace}
\newcommand{\passport}{PASSporT\xspace}
\newcommand{\passports}{\passport{s}\xspace}

\newcommand{\tmax}{t_\mathsf{max}}
\newcommand{\talive}{\epsilon_T}
\newcommand{\texp}{t_\mathsf{exp}}
\newcommand{\tnow}{\mathsf{now()}}
\newcommand{\calldt}{\mathsf{cdt}}
\newcommand{\callstr}{\src\|\dst\|\ts}

\newcommand{\ctx}{\mathtt{ctx}}

\renewcommand{\hash}{H}
\newcommand{\ok}{\mathtt{OK}}

\newcommand{\valid}{\mathsf{Valid}}
\newcommand{\audit}{\mathtt{AUDIT}}
\newcommand{\idx}{\mathsf{idx}}
\newcommand{\msg}{\mathsf{msg}}
\newcommand{\ipk}{\mathsf{ipk}}
\newcommand{\isk}{\mathsf{isk}}
\newcommand{\gsk}{\mathtt{gsk}}
\newcommand{\gpk}{\mathtt{gpk}}
\newcommand{\oprff}[2]{F_{#1}(#2)}
\newcommand{\src}{\mathsf{src}}
\newcommand{\dst}{\mathsf{dst}}
\newcommand{\ts}{\mathsf{ts}}
\newcommand{\fx}{y}
\newcommand{\fxset}{Y}
\newcommand{\filteredfxset}{\mathcal{Y}}
\newcommand{\x}{x}
\newcommand{\rand}{r}
\newcommand{\ik}{i_{k}}
\newcommand{\findms}{\mathsf{GetMS}}
\newcommand{\findevals}{\mathsf{GetEV}}
\newcommand{\findnodes}{\mathsf{GetNodes}}
\newcommand{\replparam}{m}
\newcommand{\evparam}{n}
\newcommand{\numevs}{N}
\newcommand{\nummss}{M}
\newcommand{\equal}{\stackrel{?}{=}}

\newcommand{\nodeid}{\mathsf{nid}}
\newcommand{\ipaddr}{\mathsf{nip}}
\newcommand{\nodetype}{\mathsf{ntyp}}
\newcommand{\routetable}{\mathcal{R}}

\newcommand{\retreq}{\mathtt{RETRIEVE}\text{-}\mathtt{REQ}}
\newcommand{\retres}{\mathtt{RETRIEVE}\text{-}\mathtt{RES}}

\newcommand{\EV}{EV\xspace}
\newcommand{\MS}{MS\xspace}
\newcommand{\EVs}{EVs\xspace}
\newcommand{\MSs}{MSs\xspace}

\newcommand{\SOsLong}{Admin\xspace}
\newcommand{\clearinghouse}{Clearinghouse\xspace}
\newcommand{\LS}{ALS\xspace}
\newcommand{\LSlong}{Audit Log Server\xspace}

\newcommand{\cps}{CPS\xspace}
\newcommand{\sticps}{\cps}
\newcommand{\sticr}{STI-CR\xspace}
\newcommand{\cpss}{CPSs\xspace}
\newcommand{\stipa}{STI-PA\xspace}
\newcommand{\stiga}{STI-GA\xspace}

\newcommand{\stica}{STI-CA\xspace}

\newcommand{\ca}{CA\xspace}

\newcommand{\provider}{P\xspace}

\newcommand{\setupproto}{Setup\xspace}
\newcommand{\pubproto}{Record Publish\xspace}
\newcommand{\retproto}{Record Retrieval\xspace}
\newcommand{\cidgenproto}{Call-Secret Generation\xspace}
\newcommand{\tokensproto}{Billing Token Minting\xspace}

\newcommand{\mms}{\mathtt{MS}}
\newcommand{\mmsdb}{\mathcal{D}}
\newcommand{\mev}{\mathtt{EV}}

\newcommand{\keylist}{\mathcal{K}}
\newcommand{\keylistsize}{S}
\newcommand{\rotwindow}{t_\mathsf{rot}}
\newcommand{\expidx}{i}
\newcommand{\expkey}{k_\mathsf{exp}}
\newcommand{\cskset}{\mathcal{C}}
\newcommand{\csk}{\mathsf{csk}}
\newcommand{\oprf}{\mathsf{OPRF}}
\newcommand{\oprfeval}{\mathbf{\mathsf{Eval}}}

\newcommand{\PPP}{\mathcal{P}}
\newcommand{\MMM}{\mathcal{M}}

\newcommand{\gsetup}{\mathbf{GSetup}}
\newcommand{\gkgen}{\mathbf{GKGen}}
\newcommand{\gjoin}{\mathbf{Join}}

\newcommand{\gopen}{\mathbf{Open}}
\newcommand{\ggs}{\mathcal{G}_{\mathsf{sign}}}
\newcommand{\tgs}{\mathtt{TGS}}
\newcommand{\gsign}[2]{\tgs.\sign(#1, #2)}
\newcommand{\gverify}[3]{\tgs.\verify(#1, #2, (#3))}

\newcommand{\rssign}[2]{\mathsf{RS}.\sign(#1, #2)}
\newcommand{\rsverify}[3]{\mathsf{RS}.\mathsf{Vf}(#1, #2, #3)}

\newcommand{\oobrate}{\mathcal{R}_{\text{oob}}}

\newcommand{\DDD}{\mathcal{D}}
\newcommand{\LLL}{\mathtt{L}}
\newcommand{\tkns}{\mathsf{tokens}}
\newcommand{\tkn}{\mathsf{token}}
\newcommand{\rdmd}{\mathsf{redeemed}}
\newcommand{\gettkn}{\mathtt{GET}\text{-}\mathtt{TOKEN}}
\newcommand{\checkexpiry}{\mathsf{CheckExpiry}}
\newcommand{\faulter}{\mathtt{FAULTER}}

\newcommand{\info}{\mathsf{info}}

\newcommand{\SSS}{\mathcal{S}}

\newcommand{\numtokens}{T}

\newcommand{\tks}{\mathcal{T}}

\newcommand{\foos}{\mathcal{F}_\mathsf{OOB}}
\newcommand{\register}{\mathtt{REGISTER}}
\newcommand{\genid}{\mathtt{GEN}\text{-}\mathtt{SK}} 
\newcommand{\mpub}{\mathtt{MSG}\text{-}\mathtt{PUB}}
\newcommand{\mret}{\mathtt{MSG}\text{-}\mathtt{RET}}

\newcommand{\tknlst}{\mathsf{TknList}}
\newcommand{\hyb}{\mathbf{Hybrid}}

\newcounter{takeawaycounter}
\newcommand{\takeaway}[1]{\refstepcounter{takeawaycounter}\vspace{3pt}\noindent\textbf{Takeaway~\thetakeawaycounter:}~\emph{#1}}
\newcounter{takeawayblockcounter}
\definecolor{lightgray}{RGB}{100, 100, 100}
\newmdenv[
  backgroundcolor=lightgray!15,
  linecolor=lightgray!50,
  skipabove=2pt,
  skipbelow=2pt,
  innerleftmargin=2pt,
  innerrightmargin=2pt,
  innertopmargin=2pt,
  innerbottommargin=2pt
]{takeawaybox}

\newcommand{\provReqPyld}{\mathsf{hreq}}
\newcommand{\cpsResPyld}{\mathsf{hres}}

\newcommand{\sendfeedback}[2][Log]{\vspace{2pt}{\color{gray}* #1 #2 on \LS.*}}

\newcommand{\iwf}{SIWF\xspace}

\newcommand{\SecReqLocationConfidentiality}{Record Location Confidentiality}
\newcommand{\SecReqRecordExpiry}{Record Expiry Enforcement}
 
\oobfullversionfalse

\begin{document}

\title{Secure and Efficient Out-of-band Call Metadata Transmission}
\author{\IEEEauthorblockN{David Adei}
	\IEEEauthorblockA{North Carolina State University\\
		dahmed@ncsu.edu}
	\and
	\IEEEauthorblockN{Varun Madathil}
	\IEEEauthorblockA{Yale University\\
		varun.madathil@yale.edu}
	\and
	\IEEEauthorblockN{Nithin Shyam S.}
	\IEEEauthorblockA{North Carolina State University\\
		nsounda@ncsu.edu}
	\and
	\IEEEauthorblockN{Bradley Reaves}
	\IEEEauthorblockA{North Carolina State University\\
		bgreaves@ncsu.edu}}

\maketitle

\begin{abstract}
The \fstirsha (\stirsha) attestation Framework mandated by the United States, Canada, and France to combat pervasive telephone abuse has not achieved its goals, partly because legacy non-VoIP infrastructure could not participate.
The industry solution to extend \stirsha broadcasts sensitive metadata of every non-VoIP call in plaintext to every third party required to facilitate the system. It has no mechanism to determine whether a provider's request for call data is appropriate, nor can it ensure that every copy of that call data is unavailable after its specified expiration. It threatens subscriber privacy and provider confidentiality.

In this paper, we present \sysname, a distributed, privacy-preserving system with \emph{tunable decentralization} that securely extends \fstirsha{} across all telephone network technologies. We introduce the notion of \emph{secure out-of-band signaling} for telephony and formalize its system and security requirements. We then design novel, scalable protocols that realize these requirements and prove their security within the Universal Composability framework. Finally, we demonstrate \sysname's efficiency with our open-sourced reference implementation.

Compared to the current solution, \sysname 1) protects the confidentiality of subscriber identity and provider trade secrets, 2) \emph{guarantees record expiration} as long as a single node handling a record is honest, 3) reduces resource requirements while providing virtually identical call-setup times and equivalent or better uptimes, and 4) \emph{enables secure pay-per-use billing} and integrates mechanisms to mitigate and detect misbehavior.

Moreover, \sysname can be trivially extended to provide the same security guarantees for arbitrary call metadata. Not only is \sysname a superior approach, it is also a transformative tool to retrofit fragmented global telephony and enable future improvements, such as stronger call authentication and Branded Calling.
\end{abstract} 
\IEEEpeerreviewmaketitle

\section{Introduction}\label{sec:intro}
Fraud and spam continue to plague telephone networks despite decades of mitigation, costing subscribers and providers billions each year~\cite{sipnoc24bandwidth}. Robocalls are especially pervasive, with the FCC reporting over 4 billion per month in the U.S alone. Unlike emails, illegal calls demand immediate attention, intrude on privacy, target the vulnerable, enable impersonation, harm reputations, and erode consumer trust.

Historically, regulatory approaches such as the FTC's Telemarketing Sales Rule introduced measures like the National ``Do-Not-Call'' Registry, enabling individuals to opt out of telemarketing calls. Concurrently, researchers and industry experts explored technical defenses, including allow/deny lists~\cite{pandit2018towards, tu2016sok}, reputation systems~\cite{AZAD2019841, azad2017privy}, behavioral analysis~\cite{balasubramaniyan2007callrank, mustafa2014callerdec, chu2023exploiting}, and content analysis~\cite{balasubramaniyan2010pindr0p, pandit2021applying, robohalt, sahin2017using}. These measures proved ineffective, and telephony abuse remained a threat.

Amid heightened scrutiny of telephone abuse, policymakers responded by mandating the adoption of the \fstirsha{} (\stirsha) caller attestation framework---first in the United States through the TRACED Act in 2019, and subsequently in Canada and France, with additional countries evaluating adoption. Defined in RFC 8224 \cite{rfc8224} and developed by the IETF STIR Working Group, \stirsha{} enables cryptographic verification of caller identity in SIP-based communication. Like DKIM for email, \stirsha{} requires originating providers to sign outbound calls, embedding attestation information (called \passport) in call signals for downstream providers. \stirsha also supports features such as Rich Call Data (RCD), which provides recipients with branded context to inform their call-answering decisions. 

However, like all protocols that require end-to-end delivery, \stirsha{} is fundamentally challenged by the operational reality of telephone networks, faltering in partial deployment.

\emph{\stirsha is fundamentally limited by its reliance on universal adoption to be effective.} This demands that every provider in a call path upgrade their network to support it—a task that is often infeasible on a global scale. Telephone infrastructure spans multiple generations of incompatible technologies, making such upgrades prohibitively costly and complex. Furthermore, international jurisdictional boundaries make it unrealistic for any single country to compel foreign operators to upgrade.

\emph{Second, routine signalling reconstruction at every provider gateway undermines the end-to-end integrity of \stirsha attestations.} Unlike the Internet, where packets are simply forwarded, telephone call sessions are typically terminated and re-originated to enforce internal routing policies and apply billing logic. This necessary process frequently strips \stirsha headers.

\emph{Finally, \stirsha's scope is inherently limited to IP-based networks.} It provides no native support for the vast ecosystem of legacy SS7 infrastructure that still dominates telephony. Any call that traverses a non-IP segment breaks the end-to-end chain of trust, as \passports cannot be carried over these legacy protocols.

To address the challenge of legacy interoperability, industry experts, the IETF, and the Alliance for Telecommunications Industry Solutions (ATIS)---a global standards body---approved the out-of-band \fstirsha{} (\oobss) standard~\cite{ietfoob06, atis-out-band-stir} in July 2021 to extend \stirsha{} to non-IP networks. \oobss, still in early deployment, relies on distributed third-party databases called Call Placement Services (\sticps{}), which store and disseminate \passports nationwide on behalf of providers.

However, \oobss's interoperability comes at a catastrophic cost to privacy and security. Storing sensitive call metadata—who is calling whom, when, why, and how often—in third-party databases without confidentiality creates a vast new attack surface. This grants any participating \sticps{} nationwide visibility into call patterns, meaning a single breach, compromise, or even a curious insider could enable mass surveillance, cyberattacks, and espionage. Furthermore, the protocol lacks meaningful access control, allowing any provider with a valid certificate to access unauthorized \passports. As a result, \oobss exposes the entire network's communication patterns to pervasive surveillance, undermining the privacy of subscribers and providers' trade secrets that give them a competitive edge.

This paper presents \sysname, a distributed system and protocol suite that transmits arbitrary call metadata across heterogeneous telephone networks, overcoming the limitations of \oobss{} and the technical constraints of fragmented telephony. The key insight is to establish a secure ad-hoc channel per call that is orthogonal to the existing telephone system, allowing only providers directly involved in routing the call to access its records, regardless of their peering relationships. Notably, adversaries cannot determine which \cps stores the records for a given call, nor can they link those records to the specific subscribers or providers involved. \sysname cryptographically enforces that records are permanently inaccessible after their expiry period and that they are isolated so compromising a \cps does not expose any past or future records. \sysname is modular by design, enabling each subsystem to be independently tuned along a spectrum from centralized to decentralized configurations, thus offering stakeholders deployment flexibility.

We introduce and formally define \emph{secure out-of-band signaling}, a notion that addresses the challenge of partial deployment, eliminating the need for universal adoption for telephony protocols like \stirsha. We show that \sysname meets this definition within the Universal Composability (UC) framework. Our approach reuses existing \oobss entities, imposes no new provider requirements, and integrates a cryptographic audit trail for pay-per-use billing---a more equitable billing model. \sysname also incorporates transparency mechanisms to detect misbehaving parties. Our prototype evaluation across multiple regions shows that these benefits come at a minimal performance cost, adding only a fraction of a second to call setup—roughly the blink of an eye—that is imperceptible to users.

In summary, we make the following contributions:
\begin{itemize}[itemsep=0pt, topsep=0pt, leftmargin=*]
    \item \textbf{Formalization of Out-of-Band Signaling.} We introduce secure out-of-band signaling and provide a formal definition in the Universal Composition Framework. We design corresponding protocols and prove that they meet this definition.

    \item \textbf{Privacy-preserving Metadata Handling.} We provide strong privacy for sensitive call metadata---including call unlinkability and enforced record expiry---protecting it from all parties without prior knowledge. In contrast, \oobss exposes metadata to all participants and fails to enforce its expiration.
    
    \item \textbf{Modular and Tunable Decentralization.} We provide a modular design that enables fine-grained control over trust and scalability. Each subsystem can be independently tuned along a spectrum from centralized to decentralized operation.

    \item \textbf{Resilient and Efficient Network Architecture.} Our network architecture eliminates single points of security failure and is resilient to the compromise of individual nodes. Our design is scalable—with security and performance improving as more nodes join—while reducing operational costs.

    \item \textbf{Sustainable Deployment Model.} We provide a cryptographic audit trail to enable pay-per-use billing, providing an equitable economic model for long-term operation. \sysname also includes mechanisms to detect misbehaving parties.

    \item \textbf{Open-Source Implementation and Evaluation.} We built and evaluated a prototype of \sysname across multiple AWS regions, demonstrating its efficiency and scalability with minimal overhead. We are releasing our implementation as open source, with modules to support real-world deployment.
\end{itemize}

\noindent\sysname generalizes to a distributed key-value store for sharing arbitrary metadata about live calls. This abstraction broadens its utility to support advanced defense mechanisms, which we discuss in Sec.~\ref{sec:discussion}. Ultimately, \sysname provides a robust, privacy-preserving solution to partial deployment, overcoming the challenges of universal adoption, signal reconstruction, and legacy interoperability that have hindered telephony security for decades. It thus serves as a foundation for researchers and industry stakeholders working to mitigate telephony abuse.

 \section{Background and Related Work}\label{sec:background}
Robocalls remain the most widespread form of phone abuse~\cite{usenix_whoscalling, usenix_snorcall}. In the absence of robust caller authentication~\cite{sahinsoktele, tu2016sok}, scammers can easily spoof caller IDs~\cite{mustafa2016end}, resulting in billions of dollars in losses. Existing defenses—including authentication protocols~\cite{tu2017toward, reaves2016authloop, reaves2017authenticall}, spam filtering~\cite{dantu2005detecting}, call-blocking apps~\cite{mustafa2014you, robohalt, Truecaller, du2023ucblocker}, and stricter legal penalties~\cite{fcc_tcpa_rules, tracedact}—have largely failed to deter these attacks. However, the \fstirsha attestation framework is intended to effectively prevent ``all'' caller ID spoofing attacks.

\mysubsection{\fstirsha (\stirsha) Framework}\label{sec:ss} 
\fstirsha encompasses two complementary components: the STIR protocol \cite{rfc8224,rfc8588} and the Signature-based Handling of Asserted information using toKENs (SHAKEN) specification \cite{atisshaken}. The STIR protocol standardizes the creation and use of cryptographically signed tokens embedded in call signaling. SHAKEN, developed by the ATIS/SIP Forum IP-NNI Task Force, provides implementation guidelines that ensure interoperability among service providers, translating the technical principles of STIR into a practical framework suitable for real-world telephone systems.

\fstirsha establishes a PKI to cryptographically assert authority over telephone numbers. The PKI consists of Certification Authorities (\stica{}s), who manage the certificate lifecycle and abide by policies defined by the Governance Authority (\stiga) and enforced by the Policy Administrator (\stipa). The \ca{}s issue digital certificates that bind providers to specific telephone numbers or ranges.

\begin{figure}[t]
	\centering
    \includegraphics[width=.99\columnwidth]{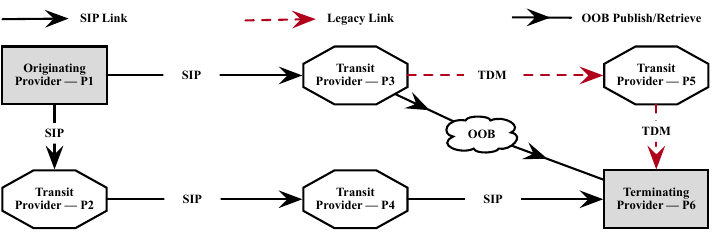}
    \caption{The telephone network spans diverse providers and signaling protocols. Because \fstirsha supports only IP-based traffic, legacy links lose attestation.}
	\label{fig:how-ss-works}
\end{figure}

\mysubsubsection{Call Routing and \fstirsha}
Fig.~\ref{fig:how-ss-works} shows a peering scenario involving six providers, highlighting their interconnection links: solid lines represent IP-based traffic using the Session Initiation Protocol (SIP), while dashed lines indicate SS7 via Time Division Multiplexing (TDM) links.

When a subscriber places a call, her provider ($\provider_1$) first verifies her authorization to use the claimed source Telephone Number (TN). If authorized, $\provider_1$ generates a digital signature over the call’s metadata in a JSON Web Token, known as a \passport, indicating the confidence level in the subscriber’s right to use the TN. Three attestation levels exist: \emph{Full (A)} — confidence in both identity and TN ownership; \emph{Partial (B)} — verified identity but uncertain TN ownership; and \emph{Gateway (C)} — confirmation only of the network entry point, with no identity or TN verification. $\provider_1$ embeds the \passport in a SIP INVITE and forwards it to the next provider, typically selected based on least-cost routing, reliability, or operational policy.

\myparagraphi{Partial Deployment Problem} In Fig.~\ref{fig:how-ss-works}, a call from $\provider_1$ to $\provider_6$ can follow one of two routes: $\provider_1 \xrightarrow{\text{\tiny SIP}} \provider_2 \xrightarrow{\text{\tiny SIP}} \provider_4 \xrightarrow{\text{\tiny SIP}} \provider_6$, or $\provider_1 \xrightarrow{\text{\tiny SIP}} \provider_3 \xrightarrow{\text{\tiny TDM}} \provider_5 \xrightarrow{\text{\tiny TDM}} \provider_6$. Regardless of the path, all providers must support \stirsha{} for the \passport{} to reach $\provider_6$; a single non-supporting provider disrupts in-band transmission. We refer to this as the \emph{partial deployment} problem. The second route highlights a challenge: $\provider_5$ operates a legacy, non-IP network. This scenario is common, as many calls—though initiated over IP—still traverse non-IP segments that cannot carry \stirsha{}. To address it, industry experts proposed out-of-band \fstirsha{} (\oobss), a mechanism that extends \fstirsha{} to non-IP networks by transmitting \passport{}s out-of-band, independent of the call signaling path via a rendezvous protocol.

\mysubsection{Out-of-Band \fstirsha (\oobss)}
Because \fstirsha is defined for SIP networks, legacy non-SIP technologies such as TDM cannot participate directly. To address this limitation, the IETF STIR working group and ATIS developed the Out-of-Band \stirsha~\cite{ietfoob06, atis-out-band-stir} (\oobss) standard, approved in July 2021, to extend \fstirsha.

Under \oobss, providers that cannot forward embedded \passport{s} in-band must publish them to a distributed network of third-party \sticps databases to which they subscribe for a fee. Each \passport is stored under the originating and destination telephone numbers. Upon publication, the receiving \sticps broadcasts (fan-out) the record to all other databases nationwide to ensure availability. When a provider receives a SIP INVITE without an embedded \passport, it must query the \sticps network using the call’s origin and destination numbers to retrieve and insert the \passport back into the call.

Consider the path $\provider_1 \xrightarrow{\text{\tiny SIP}} \provider_3 \xrightarrow{\text{\tiny TDM}} \provider_5 \xrightarrow{\text{\tiny TDM}} \provider_6$ in Fig.~\ref{fig:how-ss-works}. In this scenario, $\provider_3$ converts SIP signaling to TDM, which strips the embedded \passport. Thus, $\provider_3$ must first publish it to a \sticps before converting. $\provider_6$ then queries its subscribed \sticps instance to retrieve the corresponding \passport and sets the ``Identity'' header in the SIP INVITE accordingly.

\mysubsection{Cryptographic Primitives}

\myparagraphi{Symmetric Key Encryption} Comprises three algorithms $(\kgen, \enc, \dec)$. $\kgen$ outputs a key $\key$, $\enc(\key, \msg)$ produces ciphertext $\ctx$ and $\dec(\key, \ctx)$ returns the message $\msg$.

\myparagraphi{Verifiable Oblivious PRFs (OPRFs)} OPRFs \cite{casacuberta2022sok, jarecki2017toppss} enable a client to obtain the output $y$ of a PRF on their input $x$ without revealing $x$ to the server or learning the server's secret key. The client can further validate that $y$ is correct.

\myparagraphi{Group Signatures} Group signatures enable members to anonymously sign on behalf of the group. Traditional schemes rely on a Trusted Authority (TA) to manage group membership. Threshold Group Signature (TGS) variants \cite{bootle2016foundations,camenisch2020short} distribute the TA's role across multiple parties. We denote a TGS scheme as $\tgs = (\sign, \verify, \gpk, \gsk_i)$, where $\sign$ and $\verify$ are the signing and verification algorithms, $\gpk$ is the common group public key, and $\gsk_i$ is member $i$'s private key.
 \section{Problem Statement}

We outline the risks of metadata exposure and \oobss security limitations to motivate our work, then show how \sysname's design is guided by practical security trade-offs.

\mysubsection{It's Just Metadata, What's The Harm?}
Concerns about metadata exposure have been dismissed by statements such as ``It's just metadata, [Not the Content]''~\cite{just_metadata} Yet call metadata alone—information about whom you call, who calls you, when, and how often—can reveal deeply personal and sensitive aspects of your life: your closest relationships, health conditions, financial struggles, legal issues, reproductive choices, or identity markers such as race, religion, or sexual orientation. This risk is not theoretical; research~\cite{metadata_reveals, mayer2014metaphone} has demonstrated that call metadata alone allowed researchers to identify highly sensitive activities, such as someone battling multiple sclerosis, a woman seeking abortion services, and an individual frequently contacting a firearm dealer.

In the wrong hands, this metadata can lead to invasive profiling, targeted harassment, or worse. The risk is amplified by features like Rich Call Data (RCD), which has been proposed to carry sensitive details like name, location, and the purpose of the call. Because \stirsha lacks deniability~\cite{loomingprivdis}, its signed metadata becomes more convincing as evidence, easily misinterpreted, and severely damaging to privacy when exploited. Misuse of this information can cause significant harm~\cite{we_kill_people}.

Under \oobss{}, the risk multiplies. Call metadata is no longer restricted to the operators directly routing the call, but is shared without confidentiality across numerous third parties---all implicitly trusted with your privacy. \oobss offers no safeguards to prevent these parties from selling call patterns and phone numbers to data brokers. This creates an opaque ecosystem where a single malicious actor, compromised system, or even a curious employee can expose private details about you or anyone else targeted by adversaries.

\subsection{Security Limitations of \oobss} 
\oobss{} implicitly trusts all \sticps{}s, giving them broad visibility into nationwide call activity and making them attractive targets for adversaries from state actors~\cite{nsa_spying,nsa_collects,chinese_steals} to low-resourced individuals like intimate partners.

\myparagraphi{Lack of Confidentiality}  
\sticps{} records are stored in plaintext, leaving them vulnerable to breaches. A compromise can expose large volumes of metadata, enabling surveillance, targeted attacks, or espionage.

\myparagraphi{Subscriber Privacy Violation}
\passport{}s reveal sensitive metadata about communication patterns, timing, and social connections~\cite{heuser2017phonion}. Exposure undermines user privacy and enables both targeted and pervasive surveillance.

\myparagraphi{Trade Secret Leakage}  
\passport{}s carry provider-specific metadata that can be exploited to infer business-sensitive information, such as call volumes, peering arrangements, and traffic trends. These risks are heightened in practice when some \sticps{}s are operated by telecom providers, creating conflicts of interest and potential for misuse.

\myparagraphi{Network-wide Denial of Service} Broadcasting \passports for every call to all peers is bandwidth-intensive, difficult to scale, and creates an amplification vector exploitable to launch denial-of-service (DoS) attacks against legitimate traffic.

\subsection{\sysname's Scope and Tunable Security Trade-offs}
\myparagraphi{Scope} \sysname addresses the threats above by ensuring confidentiality for call metadata transmitted outside standard signaling paths. This prevents entities outside the call path from accessing communications, preserving privacy and enabling innovation. We clarify that \sysname is not subscriber-facing: it requires no action from users and does not require universal provider participation. \emph{However, \sysname is explicitly limited to securing out-of-band metadata. In-band voice encryption, which must account for lossy telephony codecs, remains an open challenge beyond the scope of this work.}

\myparagraphi{Security Trade-offs} Because security design is always about trade-offs, we design \sysname{} to be modular and flexible, allowing operators to tune each subsystem along a spectrum from centralized to decentralized. By adjusting the number and independence of actors, operators can balance coordination, availability, and privacy. Centralizing \sysname{} simplifies coordination and can improve availability, but creates single points of security failure and increases privacy risks. Decentralizing \sysname{} distributes trust and enhances privacy and resilience, but increases coordination overhead. This flexibility enables \sysname{} to adapt to diverse operational needs and threats.

The current \oobss protocol implicitly trusts all \cps operators with both privacy and availability. In contrast, \sysname eliminates the need to trust any single \cps; it only requires that a subset of \cpss remain semi-honest to ensure availability. This is a practical assumption, as the existing vetting processes and financial incentives within the \oobss ecosystem discourage malicious service disruption.

In the event of a service disruption, \sysname can detect and isolate misbehavior. By distributing call processing across all \cpss, a misbehaving operator cannot selectively disrupt calls---only cause random service degradation. \sysname detects misbehavior on a per-call basis, enabling rapid revocation of privileges and preserving system integrity. Crucially, under \sysname, no individual \cps can compromise subscriber privacy or provider trade secrets---not even a malicious one. 
\section{Secure Out-of-band Signaling for Telephony}\label{sec:requirements}
Consider a small coalition of providers—say, 50 of 7,300+ in the US.—who wish to deploy advanced services like branded calling or strong authentication for their customers. The current in-band, hop-by-hop model makes this impossible, as call routes change frequently, any single non-participating carrier in a call path can strip the required metadata. This creates a de facto universal adoption requirement, effectively holding innovation hostage to the most reluctant members of the ecosystem.

Intuitively, addressing partial deployment requires sidestepping non-participating providers, thus reframing the challenge as one of \emph{out-of-band (OOB) signaling}. However, this approach risks exposing sensitive data to unauthorized parties, introducing call delays, and preventing sidestepped providers from accessing necessary records. Furthermore, it can create single points of security failure and complicate troubleshooting.

An efficient OOB signaling solution that overcomes these security and reliability challenges, however, would present major opportunities. Such a mechanism would not only solve the initial problem of partial deployment but also serve as a foundation for advanced authentication and novel telephony features, unlocking a path for transformative changes in global telephony.

\subsection{System and Security Requirements}
We establish the following requirements for secure out-of-band signaling, with detailed justifications in Appendix~\ref{sec:justification} and a formal UC definition in Appendix~\ref{sec:formal-desc}.

\mysubsubsection{Functional Requirements} The system must provide core functionalities to ensure accurate message exchange.
\begin{enumerate}[label=\textbf{F\arabic*.},itemsep=0pt, topsep=2pt, leftmargin=*]
    \item\label{funcreq:f1} \underline{\emph{Record Upload and Lookup}}: The system must allow upstream providers to upload records and enable legitimate downstream providers to retrieve them and vice versa.
    
    \item\label{funcreq:f2} \underline{\emph{Correctness}}: A lookup request must return the correct record for a previously uploaded and unexpired record.

    \item\label{funcreq:f3} \underline{\emph{Efficiency}}: Record upload and lookup requests must be fast and comparable to non-secure approaches.
    
    \item\label{funcreq:f4} \underline{\emph{Scalability}}: The system must handle peak call volumes and network churn with minimal performance impact.
    
    \item\label{funcreq:f5} \underline{\emph{Resiliency}}: The system must ensure high success rates for message lookups, even when system nodes fail.
\end{enumerate}

\myparagraph{Security Requirements} No single entity --- off the call path --- should gain essential information about the call. This principle motivates the following requirements:

\begin{enumerate}[label=\textbf{S\arabic*.},itemsep=0pt, topsep=2pt,leftmargin=*]
    \item \underline{\emph{Individual Subscriber Privacy}}: Only parties with complete and accurate call details can learn that a subscriber's record exists and retrieve it.
    
    \item \underline{\emph{Call Unlinkability}}: No entity knowing the caller's number can identify recipients, and vice versa.
    
    \item \underline{\emph{Trade Secrecy}}: Only legitimate parties can retrieve their records and view aggregate data on a provider’s call volumes or peers.
    
    \item \underline{\emph{\SecReqLocationConfidentiality}}: Only parties directly participating in an active call can locate the corresponding record in the network.

    \item \underline{\emph{\SecReqRecordExpiry}}: Authorized access to records must be limited to a fixed period. Once expired, records must be irreversibly inaccessible and leak no information.

    \item \underline{\emph{Perfect Forward Secrecy}}: Even if an adversary compromises a party’s key, they must not decrypt or infer information about records exchanged prior to the compromise.
    
    \item \underline{\emph{Post-Compromise Security}}: Even after an adversary compromises a long-term key, they must not break the security of future records, assuming the system has time to recover.
\end{enumerate}

\mysubsection{Challenges of Realizing Secure Out-of-Band Signaling}
Protecting metadata outside the signaling plane creates a core dilemma: records must be accessible to all on-path providers for analytics or blocking, but kept hidden from everyone else. Satisfying both presents major architectural and cryptographic challenges, which include:

\myparagraphi{Secure Discovery} How can providers on the call path find the correct record without exposing it to the wider network? The \oobss{} broadcast model is simple but insecure, while alternatives like centralized lookup or hop-by-hop discovery~\cite{sipnoc24cpsdiscovery} avoid broadcasting but create single points of security failure and falter in partial deployment.

\myparagraphi{Multi-party Confidentiality} How to encrypt records so that only on-path providers can access them. This is challenging because each provider knows only its immediate neighbors, not the full call path. A common symmetric key shared by providers is insecure, as a single provider's compromise would expose all records. Standard Public-key encryption is similarly problematic: encrypting for the final destination blinds intermediary providers who need access, while encrypting for a set of public keys is impractical, as the encryptor cannot predict the full call path in advance.

\myparagraphi{Enforced Data Ephemerality} Outsourcing call metadata to third parties creates a historical archive of growing liability. \oobss requires operators to delete the records after 15 seconds, but provides no cryptographic mechanism to enforce it. The core challenge is how to cryptographically enforce data expiry, ensuring records become permanently inaccessible after a fixed period, even if the nodes are malicious or compromised.

\section{Our Approach to Out-of-Band Signaling}
We present the \sysname{} system that realizes secure out-of-band signaling, then discuss security tuning, pay-per-use billing, misbehavior detection, and address common questions.

\mysubsection{Base \sysname System}\label{sec:base-sys}

\begin{center}
    \resizebox{.99\columnwidth}{!}{\newlength\BaseSysBoxWidth
\setlength\BaseSysBoxWidth{3.2cm}
\newlength\BaseSysBoxHeight
\setlength\BaseSysBoxHeight{0.8cm}
\newlength\ProviderSep
\setlength\ProviderSep{9cm}

\begin{tikzpicture}[>=stealth, thick]
\node[draw, rounded corners, fill=blue!10,
        minimum width=\BaseSysBoxWidth,
        minimum height=\BaseSysBoxHeight,
        anchor=east] (admin) at (5, 1) {Administration};
  \node[draw, rounded corners, fill=orange!15,
        minimum width=\BaseSysBoxWidth,
        minimum height=\BaseSysBoxHeight,
        anchor=east] (ev)    at (5, 0) {Evaluator (\EV)};
  \node[draw, rounded corners, fill=red!10,
        minimum width=\BaseSysBoxWidth,
        minimum height=\BaseSysBoxHeight,
        anchor=east] (ms)    at (5, -1) {Message Store (\MS)};

\path (admin) -- (ms) coordinate[pos=0.5] (mid);

\node[draw, rounded corners, fill=green!10,
        minimum width=\BaseSysBoxWidth,
        minimum height=\BaseSysBoxHeight]
    (provider) at ($(mid)+(-\ProviderSep,0)$) {Provider $(\provider_i)$};

\coordinate (tmpAdmin) at ($(provider.east)+(1cm,1cm)$);
  \coordinate (tmpMS)    at ($(provider.east)+(1cm,-1cm)$);

\draw[->]
    (provider.east)
      -- (tmpAdmin)
      -- node[midway, above] {Acquire \sysname access}
    (admin.west);

  \draw[->]
    (provider.east)
      -- node[midway, above] {Derive ephemeral keys}
    (ev.west);

  \draw[->]
    (provider.east)
      -- (tmpMS)
      -- node[midway, above] {Share encrypted metadata}
    (ms.west);
\end{tikzpicture}
     }
    
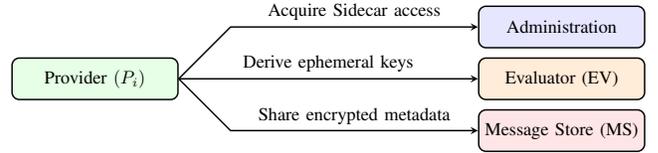
\captionof{figure}{\sysname comprises three subsystems: Administration (\SOsLong) manages membership and registers \cps nodes and providers; Evaluator (\EV) derives shared encryption keys with providers; and Message Store (\MS) caches encrypted records for up to $\tmax$ seconds.}
    \label{fig:base-sys}
\end{center}

\mysubsubsection{Strawman Out-of-Band Signaling}
Fig.~\ref{fig:base-sys} illustrates the basic \sysname setup. Although a single trusted party could manage all three subsystems, we assume the \stirsha Policy Admin manages the \SOsLong subsystem, while one designated \cps manages the \EV subsystem, and another \cps manages the \MS subsystem.

In the current \stirsha{} ecosystem, the \SOsLong enforces policies established by the \stirsha governing body, including the authorization of service providers, and \cps operators. This remains unchanged in \sysname. To participate, a provider $\provider_i$ completes a one-time registration with the \SOsLong to obtain \sysname access---the capability to authenticate to \cps nodes.

Recall that a single call may traverse multiple provider hops. The providers along this path form a message channel and can exchange messages as needed. To share a \passport $\msg$, a provider $\provider_i$ interacts with the \EV to derive a shared ephemeral secret key $\csk$ unique to the call, encrypts $\msg$ as $\ctx$, computes $\idx$, and sends $(\idx, \ctx, \sigma)$ to the \MS, where $\sigma$ is a signature verifiable with the \SOsLong's public key. Any other provider~$\provider_j$ on the call path can compute $\csk$ and $\idx$, retrieve $\ctx$ from the \MS, and decrypt it to obtain $\msg$.

The remainder of this section details the cryptography and design choices needed to secure the strawman solution.

\mysubsubsection{Toward Confidential Metadata}
Providing confidentiality for a call metadata raises a critical question: which providers should access the shared messages? Ideally, access is limited to those directly responsible for routing the call—that is, providers on the call path. However, due to the signal reconstruction at gateways, there is no unique identifier to determine the precise set of on-path providers. In practice, only the source, i.e, Caller ID ($\src$) and destination ($\dst$) telephone numbers are known to the providers on the call path. 

To achieve confidentiality, a natural approach is to encrypt $\msg$ under the public keys of all providers on the call path. However, this is impractical, as call routes change frequently and the encryptor does not know all downstream hops. Using the tuple $\calldt = (\src, \dst)$ to derive a call secret key $\csk \gets H(\calldt)$ is also insecure due to its low entropy—any party can easily decrypt records for calls they can guess. Adding the call timestamp $\ts$ (rounded to the nearest minute for instance) and using $\calldt = (\src, \dst, \ts)$ helps, since $\ts$ causes avalanche effect on $H(\calldt)$, differing for each call rather than being constant for each pair $(\src, \dst)$. Nevertheless, this construction alone is still unsuitable as the shared secret key for encryption.

To address this, an initial approach is to compute the call secret $\csk \gets \oprff{\sk}{\calldt}$ using a keyed pseudorandom function (PRF) with a shared secret key, $\sk$. However, sharing $\sk$ among all providers is insecure, as a single compromise would expose the key and break privacy. A more secure alternative is to let the Evaluator (\EV) hold $\sk$ and perform the PRF computation on the providers' behalf. Once again, this approach introduces its own privacy leak, as it reveals the call tuple $(\src, \dst, \ts)$ to the Evaluator who needs it to compute $\oprff{\sk}{\calldt}$.

We adapt the Oblivious PRF (OPRF) protocol to compute \(\oprff{\sk}{\calldt}\) without revealing either $\src$, $\dst$, or $\ts$ to the \EV. The provider \(\provider_i\) computes \(\x \leftarrow H_1(\calldt)^\rand\) which blinds $\calldt$ using a random exponent \(\rand\), and sends \(\x\) to the \EV, who responds with \(\fx \leftarrow \oprff{\sk}{\x}\). Here, $\x$ and $\fx$ are elements of the elliptic curve group $G$. The provider computes \(\csk = \fx^{1/\rand}\in G\), and verifies that \(\csk\) was computed honestly by using the public key \(\pk\) corresponding to the \(\sk\) used by the \EV. \emph{At the end of this interaction, the \EV neither learns $\calldt$ nor the secret $\csk$ itself while $\provider_i$ learns only $\csk$ but not \EV's secret $\sk$.}

The constructed $\csk$ is known only to legitimate hops on that call. To encrypt $\msg$, $\provider_i$ samples a $\lambda$-bit random string $c_0$, computes $c_1 \leftarrow \enc(\hash(c_0 \| \csk), \msg)$, and sets $\idx \gets H(\csk)$, $\ctx\gets (c_0, c_1)$. It then sends $(\idx, \ctx, \sigma)$ to the \MS. Upon receiving the request, the \MS verifies $\sigma$ and caches the record for $\tmax$ seconds, after which the record expires and is no longer retrievable. The retrieving provider $\provider_j$ can independently compute $\csk$, retrieve the record, and decrypt it.

\mysubsubsection{Record Expiration and Resilience to Compromise}
Using $\csk$ as the encryption key ties confidentiality to the secrecy of the \EV's secret key $\sk$. If the \EV{} is compromised or even just curious, confidentiality is lost for all past and future communications. Additionally, the \MS stores records indefinitely. How can we cryptographically enforce that records become unusable after their specified expiration?

\emph{We address both issues with a novel key rotation protocol that gives \sysname{} Perfect Forward Secrecy (so past messages remain secure), Post-Compromise Security (so future messages stay protected), and automatic record expiry even if the \EV's $\sk$ is leaked.} The \EV{} maintains a list $\keylist$ of $\keylistsize$ keys, cycling through them by replacing $\keylist[\expidx]$ with a freshly sampled $(\sk_\expidx, \pk_\expidx)$ every $\rotwindow$ seconds and incrementing $\expidx = (\expidx + 1) \bmod \keylistsize$. Thus, key access at any time reveals nothing about other epochs. 

During the OPRF protocol, a provider computes an index \(\ik = \hash(\calldt) \bmod \keylistsize\) which tells the \EV which key to use and includes it in the request. The \EV uses \(\keylist[\ik]\) (i.e $(\sk, \pk)$ key pair at index $\ik$) to return \((\fx, \pk)\) to the provider. Here, \(\fx\) is computed with \(\sk\) and $\csk$ (i.e $\fx^{1/\rand}$) verifies against \(\pk\).

Rarely, the $\ik$-th key in $\keylist$ may rotate just before the retrieving provider runs the OPRF, causing a mismatch with the $\csk$ used to publish the \passport{}. To handle this edge case, the \EV{} returns two outputs: $(\fx, \pk)$ from the current key, and $(\fx', \pk')$ from the just-expired key if $\ik = (\expidx-1)\bmod\keylistsize$ and the rotation occurred within the last $\talive$ seconds (small); otherwise, it returns only $(\fx, \pk)$. The publishing provider uses $\csk$ from $\fx$, while the retriever tries both $(\csk, \csk')$ for retrieval.

\mysubsubsection{Towards Authenticity and Integrity}
Although providers hold \stirsha credentials verifiable up to the Root CA, directly authenticating to the \EV{} or \MS{} exposes their communication patterns to traffic analysis. For example, a curious \cps could track aggregate call volumes or even infer the peering relationships of any provider they monitor.

To address this, \emph{we use an anonymous signature scheme known as group signatures. Group signatures allow authorized members to sign messages on behalf of the group without revealing their identity, while still enabling traceability and revocation of misbehaving members.} Signatures are verified using a common group public key $\gpk$, though each member $i$ holds a unique secret signing key $\gsk_i$. Each provider $\provider_i$ obtains $\gsk_i$ from the \SOsLong{} and uses it to sign all messages sent to the \EV{} and \MS{}, which reject any unsigned or invalidly signed communication under $\gpk$.

\myparagraphi{Access Revocation}
The \SOsLong publishes an Access Revocation List, which the \EV and \MS maintain a copy of and query to decide whether to reject a publish or retrieve request.

\mysubsection{Tunable Decentralization and Trust Distribution}
We show how \sysname distributes trust using $t$-of-$n$ threshold cryptography, enabling tunable decentralization. This approach eliminates single points of security failure in the semi-centralized base system (Sec.~\ref{sec:base-sys}), so privacy and availability guarantees depend on a threshold of entities, not just one. While distributing trust strengthens security, it introduces coordination overhead and the challenge of ensuring providers independently select the same set of \cps nodes in the \EV or \MS subsystems for a given call. Our goals for distributing trust are:
\begin{itemize}[itemsep=0pt, topsep=2pt, leftmargin=*]
    \item \underline{\emph{Security Tuning}}: Provide a ``knob'' to flexibly and independently tune \sysname's \SOsLong, \EV, and \MS subsystems along a spectrum from centralized to decentralized configurations, enabling trade-offs between privacy, availability, and delay.
    \item \underline{\emph{Consistent Content Addressing}}: Enable providers on the same call path to independently and consistently select the same set of \cps{} operators for coordination.
    \item \underline{\emph{Cryptographic Load Balancing}}: Distribute requests uniformly across \cps{} operators to minimize targeted disruption and ensure robust system performance.
\end{itemize}

\mysubsubsection{Decentralizing Administration Subsystem}
The \SOsLong{} in the base \sysname{} system functions as a $1$-of-$1$ scheme. We generalize this to an $a_I, a_O$-of-$A$ threshold model, adapting threshold Group Signature (TGS)~\cite{bootle2016foundations,camenisch2020short} variants to manage group membership. In this model, a provider must interact with a threshold of $a_I\geq1$ out of $A$ \SOsLong{} entities to join \sysname, while a threshold of $a_O\geq1$ must cooperate to deanonymize a signature. \emph{This design is critical as it ensures no single \SOsLong{} can unilaterally deny service to a new provider or maliciously deanonymize an honest one.}

\mysubsubsection{Decentralizing Message Store Subsystem}
The \MS{} in the base \sysname{} operates as a $1$-of-$1$ store, which we generalize to a $\replparam$-of-$\nummss$ scheme using a \emph{replication parameter} $\replparam \geq 1$ for redundancy and improved availability, and $\nummss$ is the total number of \MS nodes. In this model, providers replicate records across $\replparam$ message stores and can parallelize both publishing and retrieval requests to minimize latency.

A key challenge is ensuring all on-path providers independently select the same $\replparam$ out of $\nummss$ nodes for each call. \emph{We address this by modeling the authoritative message stores as a function of the call secret $\csk$, consequently hiding record locations from unauthorized parties.} Specifically, we use the XOR distance metric from Kademlia~\cite{maymounkov2002kademlia}, where the distance between two values is their bitwise exclusive OR (XOR). The $\replparam$ nodes whose unique IDs are closest to $\csk$ by this metric are selected as authoritative. We denote this selection as $\{\mms_1, \dots, \mms_{\replparam}\} = \findms(\csk, \replparam)$. \emph{Because $\csk$ is random for each call, the set of authoritative \MS{} nodes shifts frequently, providing natural load balancing across the subsystem.} We discuss more on content addressing in Appendix~\ref{sec:content-addr}.

\mysubsubsection{Decentralizing Evaluator Subsystem}
The Evaluator can be generalized from a $1$-of-$1$ setup to a $\evparam$-of-$\numevs$ scheme using an \emph{evaluation parameter}, $\evparam$, and the total number of \EV nodes $\numevs$. Unlike with \MSs, the authoritative \EVs{} must be selected before the call secret $\csk$ is generated. Selection is therefore based on a hash of the call details, $\calldt \gets H(\callstr)$. A provider computes $\{\mev_1, \dots, \mev_{\evparam}\} \gets \findevals(\calldt, \evparam)$ by selecting the $\evparam$ \EVs{} whose IDs are closest to $\calldt$ using the XOR distance metric. The provider then runs the OPRF protocol with this set of \EVs{} to obtain group elements $(\fx_1, \ldots, \fx_\evparam)$ and computes element $Y\gets (\prod_{j=1}^{\evparam} \fx_j) \in G$ and the final secret $\csk \gets H(Y^{1/\rand})$, where $\rand$ is the random blinding exponent.

\emph{This construction guarantees forward secrecy, post-compromise security, and record expiration as long as at least one of the $\evparam$ Evaluators follows the key rotation protocol, despite the curiosity of the remaining $(\evparam-1)$ \EVs. Note that \EVs{} rotate keys independently and do not synchronize clocks.}

\mysubsection{Cryptographic Audit Trail for Pay-per-Use Billing}
Telephony is inherently revenue-driven, so a platform like ours must consider billing. Fixed subscription billing can unfairly allocate costs and revenue. A pay-per-use model is more equitable but requires verifiable usage tracking. \emph{\sysname integrates a cryptographic audit trail based on tokens as an optional component to support pay-per-use (PPU) billing, so stakeholders can adopt it if need be.} Our goals for PPU are:
\begin{enumerate}[itemsep=0pt, topsep=2pt, leftmargin=*]
    \item \underline{\emph{Provider costs}} scale with the number of calls requiring out-of-band signaling service.
    \item \underline{\emph{\cps{} operator revenue}} reflects their availability and the proportion of calls they serve.
    \item \underline{\emph{Tamper-evident audit trails}} enable detection of double spending and ensure fair compensation.
\end{enumerate}

\noindent\myparagraphi{Token Life-cycle}
Our pay-per-use model uses tokens with pre-agreed values and billing cycles. At the start of each cycle, providers purchase tokens from the clearinghouse (\SOsLong) via a Verifiable OPRF interaction, during which the clearinghouse learns nothing beyond the number of tokens issued. This interaction creates tokens that are \emph{anonymous yet linkable} for accountability, \emph{tied to specific} \cps operators, \emph{valid only for the current cycle}, and \emph{redeemable only upon use}. Each out-of-band call consumes a token, which is shared among servicing operators who then redeem it with the clearinghouse.

\myparagraphi{Reconciliation (Automated)}
At the end of each cycle, the clearinghouse reconciles accounts, detects double-spending, and deanonymizes conflicts to ensure accountability. Billing keys are refreshed each cycle, making tokens single-use per cycle. Unused tokens can be returned for rollover credit.

\subsection{Detecting Misbehaving Parties}
No system can realistically prevent all forms of misbehavior. Instead, secure systems must detect misbehavior and minimize its impact. While, in practice, \cps operators are rigorously vetted and share a financial incentive for service correctness, compromises can still occur, whether intentional or not. \emph{By detecting misbehavior, \sysname enables operators to identify compromises, respond appropriately, and, if necessary, revoke access for malicious or faulty \cps nodes.}

\mysubsubsection{Transparency Mechanisms}
\sysname monitors and audits \cps operators via a public registry $\routetable$ and a centralized, append-only log \LSlong (\LS), both managed by the \SOsLong.

\myparagraphi{\sticps Pulse}
The public \cps registry \(\routetable\) node periodically sends heartbeats to each \sticps, logging metrics such as uptime and availability. These metrics are later used to assess compliance with \sticps Service Level Agreements (SLAs).

\myparagraphi{Key Rotation Transparency}
Recall that \EVs rotate their keys every $\rotwindow$ seconds. Each \EV{} must publish every rotated public key, along with its index, timestamp, and a signature, to the append-only \LS. The \LS verifies and records these entries for future auditing, enabling auditors to confirm correct key rotations and correlate outputs with provider reports.

\myparagraphi{Protocol Feedback Logging}  
Each \cps (both \EVs and \MSs) logs every protocol interaction to the \LSlong, while providers log only upon detecting misbehavior, such as invalid signatures or inconsistent $\csk$. However, providers cannot frame honest \cps operators. These verifiable logs enable token aggregation, \cps payment, and detection of misbehavior. Although logs are submitted in real time, they are latency-insensitive and must run in the background.

\mysubsubsection{Mitigating and Detecting Malicious Evaluators}
Decentralizing the \EV{} subsystem introduces a direct trade-off between resilience and a new availability risk. While computing the key $\csk$ from multiple OPRF outputs prevents a single point of failure for confidentiality, privacy and record expiry, it also exposes the system to service disruptions from any of the individual evaluators. A malicious \EV{} can return inconsistent outputs and randomly disrupt the fraction of calls processed by it. This section details how \sysname manages this trade-off through both mitigation and detection strategies.

\noindent\myparagraphi{Mitigation Strategies}
\sysname's tunable architecture provides several complementary strategies to mitigate this risk.
\begin{itemize}[itemsep=0pt, topsep=0pt,leftmargin=*]
    \item \underline{\em Restricted \EV Decentralization}: The \EV{} subsystem can be restricted to a small, fixed set of trusted operators, for example, by deploying a $2$- or $3$-of-$5$ scheme for $\evparam\geq1$ and $\numevs=5$. This configuration provides the necessary resilience against a single point of security failure while simplifying the trust model to a small, fixed set of operators.
    \item \underline{\em Increased Entry Barrier}: Operators can implement strict onboarding processes to raise entry barriers,  ensuring only thoroughly screened parties serve as \EVs.
    \item \underline{\em Economic Deterrents}: \sysname stakeholders can impose meaningful penalties for malicious behavior upon detection, establishing strong deterrents against misbehavior.
\end{itemize}

\noindent\myparagraphi{Detection} \sysname detects this misbehavior through:
\begin{itemize}[itemsep=0pt, topsep=0pt,leftmargin=*]
    \item \underline{\em Key Rotation Auditing}: The \LS{} analyzes log patterns for key rotation anomalies, such as unusual timing or unexpected duration of public keys at a given index.
    \item \underline{\em Provider Feedback}: A provider's report flags evaluator $\mev_j$ as ``likely dishonest'' if the response $(\fx, \pk_{\ik}, \sigma_r)$ it receives fails any of three cryptographic checks: (a) $\sigma_r$ fails verification under the \EV's long-term key (b) the computed $\csk$ fails verification under $\pk_{\ik}$, or (c) $\pk_{\ik}$ does not match the \LS{} log at index $\ik$ for the relevant time period.
    \item \underline{\em SLA Monitoring}: The metrics in $\routetable$ enable determining the \EV's SLA compliance to detect \EVs maliciously down.
\end{itemize}

\subsection{Frequently Asked Questions}

\myquestion{Does \sysname authenticate subscribers, and how does it handle roaming subscribers}
No, \sysname is not an authentication system and does not interact directly with subscribers. Instead, it conveys authentication-related metadata to support telephony security protocols that require end-to-end delivery. \sysname is location-agnostic: cryptographic operations depend on provider keys, not the roaming subscriber's location. As with \oobss, any provider along the call path—not just the originating provider—can interact with \sysname.

\myquestion{Does \sysname modify legacy interfaces} \sysname deployment is restricted to the gateways handling protocol translations, where call metadata would otherwise be lost, eliminating the need to directly modify legacy systems. We have developed plugins that augments these gateways with \sysname capabilities.

\myquestion{Wouldn’t encrypting call metadata hinder Lawful Interception}  
Lawful Interception (LI) systems operate independently of \fstirsha, \oobss, and \sysname, which make no changes to LI infrastructure. \sysname only protects metadata from third parties, without encrypting Call Detail Records or audio. Providers can still fully support authorized LI requests.

\myquestion{Who can operate an \EV, an \MS, or \SOsLong} 
In a deployment intended to replace \oobss, only authorized \cpss would serve as \EV or \MS nodes, and the \stirsha governing body would operate the \SOsLong. For new services such as branded calling, the coalition of providers could establish their own rules for who operates each role.

\myquestion{What if some providers refuse to participate in \sysname}
\sysname does not require all providers to participate to provide value. For any given call, the security benefits are realized by the on-path providers who participate. It serves as an interoperable replacement for the privacy-invasive \oobss protocol, and we anticipate its adoption will be driven by industry demand for better security.

\myquestion{What if regulatory mandates refuse to adopt \sysname}
\sysname's design as a platform for innovation creates a direct business incentive for any coalition of providers to deploy it for new, revenue-generating services like branded calling, offering a return on investment beyond simple compliance.

\myquestion{Isn't this just a transitional problem solvable by upgrading telephony equipment}
The telephone network is nearly 150 years old, and its legacy SS7 infrastructure still dominates global telephony. Universal upgrades are infeasible due to cost, technical complexity, and international barriers—no single country can realistically compel foreign operators to upgrade. If upgrading were simple, standards bodies would not invest in \oobss, which remains in early deployment. The network's persistent reliance on legacy systems is not a temporary issue but a structural reality, making backward-compatible and incrementally deployable solutions necessary.

\myquestion{How will \sysname remain relevant as providers transition to all-IP}
\sysname addresses partial deployment in telephony. Even as networks migrate to all-IP, hop-by-hop signaling, signal reconstruction, and universal adoption persist and threaten end-to-end security. \sysname solves these problems, so deploying mechanisms like \stirsha would require participation from only customer-facing providers.
  \section{System and Protocol Specification}\label{sec:protocolspecs}
\noindent We present \sysname's system architecture, threat model, protocol specification, and security proof.

\begin{figure}[t]
    \centering
    \includegraphics[width=.99\columnwidth]{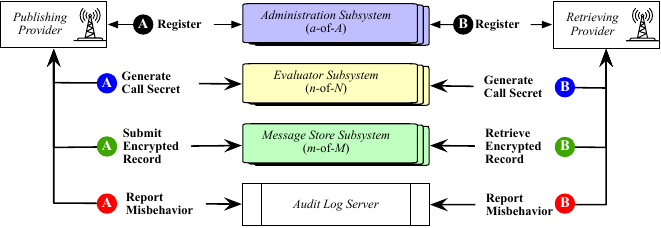}
    \caption{Call flow from $\provider_i$ to $\provider_j$ using \sysname.  Both providers register once with the \SOsLong to obtain access (\protect\circletext{black}{A}, \protect\circletext{black}{B}). $\provider_i$ interacts with $\evparam \ll \numevs$ \EVs to derive a shared call secret (\protect\circletext{blue}{A}), then replicates the ciphertext across $\replparam \ll \nummss$ \MSs with anonymous authentication (\protect\circletext{green}{A}). $\provider_j$ independently computes the call secret (\protect\circletext{blue}{B}) and accesses the record (\protect\circletext{green}{B}). Either provider can report suspected misbehavior (\protect\circletext{red}{A}, \protect\circletext{red}{B}).}
\label{fig:architecture}
\end{figure}

\subsection{System Actors, Threat Model and Protocol Overview}
\myparagraph{Actors} Fig.~\ref{fig:architecture} depicts a typical operation of \sysname. A \underline{\emph{Provider}} may publish or retrieve metadata about a live call identified by $(\src,\dst,\ts)$. A \underline{\emph{Message Store}} caches and retains encrypted messages for up to $\tmax$ seconds. An \underline{\emph{Evaluator}} computes $\oprff{\sk}{\x}$ on provider-supplied input $\x$ and returns the result. The \underline{\emph{\SOsLong{s}}} manage group membership and revoke misbehaving providers and \cps operators. They also operate the public registry $\routetable$ and the centralized \LSlong (\LS).

\mysubsubsection{Threat Model}
We assume all \sysname system actors (e.g., \cps, \SOsLong, providers) and external actors (e.g., private investigators, intelligence agencies, identity thieves) are fully malicious with respect to privacy, seeking to extract metadata or disrupt the system. For availability, we assume a majority of \EVs are semi-honest.
We allow collusion among any subset of providers, message stores, and evaluators. We show that \sysname maintains its security guarantees under these adversarial conditions.

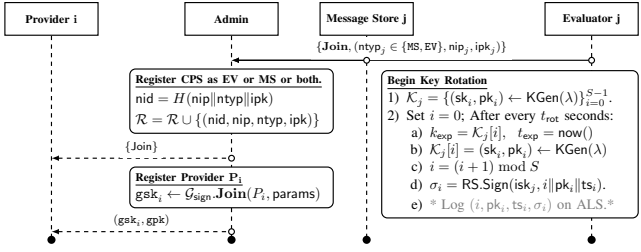
\begin{figure}[t]
    \centering
    \resizebox{.99\columnwidth}{!}{\newlength\SetupProtoBoxSpacing
\setlength\SetupProtoBoxSpacing{0.8cm}
\newlength\setupProtoLineHeight
\setlength\setupProtoLineHeight{-5cm}
{
\begin{tikzpicture}[node distance=1cm and 2cm]
\node[actor, anchor=north west, text width=2cm] (Pi) at (0,0)
    {\textbf{Provider} $\mathbf{i}$};
  \draw[dashed,thick] (Pi.south) -- ++(0,\setupProtoLineHeight) node[lifeline] {};

\node[actor, text width=2cm, right=2.1cm of Pi] (SOs) {\textbf{\SOsLong}};
  \draw[dashed,thick] (SOs.south) -- ++(0,\setupProtoLineHeight) node[lifeline] {};
  
  \node[protobox, text width=4.7cm, below=\SetupProtoBoxSpacing of SOs.south] (SOsB2) {
    \protoboxtitle{Register \sticps~as~\EV~or~\MS~or both.}
    {\setstretch{1.2}
    $\nodeid = \hash(\ipaddr \|\mathsf{ntyp} \|\ipk)$\par 
    $\routetable = \routetable \cup \{(\nodeid,\ipaddr,\mathsf{ntyp},\ipk)\}$
    
    }
  };
  
  \node[protobox, text width=4.7cm, below=0.8cm of SOsB2.south] (SOsB3) {
    \protoboxtitle{Register Provider $\mathbf{\provider_i}$} 
    $\gsk_i \leftarrow \ggs.\gjoin(P_i,\mathsf{params})$
  };

\node[actor, text width=2cm, right=1cm of SOs] (MSj)
    {\textbf{Message Store j}};
  \draw[dashed,thick] (MSj.south) -- ++(0,\setupProtoLineHeight) node[lifeline] {};

\node[actor, anchor=north east, text width=2cm, right=3.2cm of MSj] (EVj)
    {\textbf{Evaluator j}};
  \draw[dashed,thick] (EVj.south) -- ++(0,\setupProtoLineHeight) node[lifeline] {};

  \node[protobox, anchor=north east, text width=6cm]
    (EVjB2) at ($(EVj.south east)+(0,-0.84cm)$)
  {
    \protoboxtitle{Begin Key Rotation}

    {
    \begin{enumerate}[itemsep=0pt, topsep=0pt,leftmargin=*]
        \item $\keylist_j=\{(\sk_i, \pk_i) \leftarrow\kgen(\lambda)\}_{i=0}^{\keylistsize-1}$.
        \item Set $\expidx=0$; After every $\rotwindow$ seconds:
            \begin{enumerate}[itemsep=0pt, topsep=0pt,leftmargin=*]
                \item $\expkey = \keylist_j[\expidx],~~\texp = \tnow$
                \item $\keylist_j[\expidx]=(\sk_\expidx,\pk_\expidx)\leftarrow\kgen(\lambda)$
                \item $\expidx=(\expidx+1)\bmod\keylistsize$
                \item $\sigma_{\expidx}=\rssign{\isk_j}{\expidx\|\pk_\expidx\|\ts_\expidx}$.
                \item \sendfeedback{$(\expidx,\pk_\expidx,\ts_\expidx,\sigma_\expidx)$}
            \end{enumerate}
    \end{enumerate}
    }
  };

  \draw[dashed,<-,>=latex,thick,msg res]
    ($(Pi.south)+(0,-3cm)$) --
    ($(SOs.south)+(0,-3cm)$)
    node[pos=0.5,above,fill=white] {
    \footnotesize
    $\{\mathsf{Join}\}$
    };
    
  \draw[dashed,<-,>=latex,thick,msg res]
    ($(Pi.south)+(0,-4.8cm)$) --
    ($(SOs.south)+(0,-4.8cm)$)
    node[pos=0.5,above,fill=white] {
    \footnotesize
    $(\gsk_i,\gpk)$
    };

  \draw[->,>=latex,thick,msg req]
    ($(EVj.south)+(0,-0.6cm)$) --
    ($(SOs.south)+(0,-0.6cm)$)
    node[pos=0.5,above,fill=white] {
    \footnotesize
      $\{\mathbf{Join},(\mathsf{ntyp}_j \in \{\mms, \mev\},\ipaddr_j,\ipk_j)\}$
    };

\draw[->,>=latex,thick,msg req]
    ($(MSj.south)+(0,-0.6cm)$) --
    ($(SOs.south)+(0,-0.6cm)$);

\end{tikzpicture}
}
 }
    \caption{\textbf{\emph{\setupproto Protocol}:} The \SOsLong initializes the group by running $(\gpk, \mathsf{params}) \gets \gsetup(1^\lambda)$. Fig.~\ref{fig:protosetup} illustrates the protocol interactions. Each \EV initiates the key rotation protocol and logs corresponding $\pk_\expidx$ to the \LS. Providers register with the \SOsLong to obtain system credentials $(\gsk_i, \gpk)$.}
	\label{fig:protosetup}
\end{figure}

\begin{figure}[t]
    \centering
    \resizebox{.99\columnwidth}{!}{\newlength\BillingProtoBoxSpacing
\setlength\BillingProtoBoxSpacing{0.1cm}
\newlength\BillingProtoLineHeight
\setlength\BillingProtoLineHeight{-7.0cm}

{
\begin{tikzpicture}[node distance=1cm and 2cm]
\node[actor, anchor=north west, text width=2cm] (Pi) at (0,0) {\textbf{Provider} $\mathbf{i}$};
  \draw[dashed,thick] (Pi.south) -- ++(0,\BillingProtoLineHeight) node[lifeline] {};
  \node[protobox, anchor=north west, text width=9.4cm] (PiB1) at ($(Pi.south west)+(0,-\BillingProtoBoxSpacing)$) {
    \protoboxtitle{B.~Compute Tokens}
    For each $j \in [\numtokens_i]$, do:
        \begin{enumerate}[itemsep=0pt, topsep=0pt]
            \item $a_j \leftarrow \{0, 1\}^{\lambda}$, $t_{j,0} = \hash(P_i\|a_j)$,  $r_j \leftarrow \ZZ_q$, $x_j = H_1(t_{j,0})^{r_j}$
        \end{enumerate}
    Sign payload: $\sigma = \rssign{\isk_i}{x_1\|...\|x_{\numtokens_i}}$
  };
  \node[protobox, text width=8.5cm, anchor=north west,] (PiB2) at ($(PiB1.south west)+(0,-2.4cm)$) {
    \protoboxtitle{D.~Retrieve Tokens}
    {
        For $j \in [\numtokens_i]$, do:
        \begin{enumerate}[itemsep=0pt, topsep=0pt, leftmargin=*]
            \item $t_{j,1} = (\fx_j)^{1/r_j}$ and skip if $e(\vk_b, H_1(t_{j,0})) \neq e(g, t_{j,1})$
            \item Add token to list: $\tks_i = \tks_i \cup \{(t_{j,0}, t_{j,1})\}$.
        \end{enumerate} 
        Output $\tks_i$
    }
  };
  
\node[actor, anchor=north east, text width=2cm] (CH) at (13.6cm,0cm) {\textbf{\SOsLong}};
  \draw[dashed,thick] (CH.south) -- ++(0,\BillingProtoLineHeight) node[lifeline] {};
  
  \node[protobox, anchor=north east, text width=3.4cm] (CHB1) at ($(CH.south east)+(0,-\BillingProtoBoxSpacing)$) {
    \protoboxtitle{A.~Init Billing Cycle}
    $(\sk_b, \vk_b) \leftarrow \kgen(\lambda)$
  };
  
  \node[protobox, anchor=north east, text width=6.5cm] (CHB2) at ($(CHB1.south east)+(0,-1.5cm)$) {
    \protoboxtitle{C.~Endorse Tokens} 
    \begin{enumerate}[itemsep=0pt, topsep=0pt, leftmargin=*]
        \item Abort if $\rsverify{\ipk_i}{\sigma}{x_1\|...\|x_{\numtokens_i}} \neq 1$.
        \item $y_j = (x_j)^{\sk_b} ~\forall j \in  [\numtokens_i]$
    \end{enumerate}
  };

\draw[->, >=latex, thick, msg req]
    ($(Pi.south)+(0,-2.5cm)$) 
    -- 
    ($(CH.south)+(0,-2.5cm)$) 
    node[pos=0.5, above,fill=white] {
    \footnotesize
        $\{\oprfeval, (\{x_j~\forall j \in [\numtokens_i]\}, \sigma)\}$
    };
    
\draw[dashed, <-, >=latex, thick, msg res]
    ($(Pi.south)+(0,-4.2cm)$) 
    -- 
    ($(CH.south)+(0,-4.2cm)$) 
    node[pos=0.25, above,fill=white] {
        $(\{y_j\mid\forall j \in [\numtokens_i]\})$
    };
\end{tikzpicture}
} }
    \caption{\textbf{\emph{\tokensproto Protocol}:} Each billing cycle uses a fresh key pair $(\sk_b, \vk_b)$ for token verification. A provider obtains a batch of tokens from the \SOsLong via VOPRF. Providers generate identity-bound pre-tokens, which the \SOsLong endorses. $\provider_i$ verifies each token under $\vk_b$ and stores them for future.}
	\label{fig:tokens}
\end{figure}
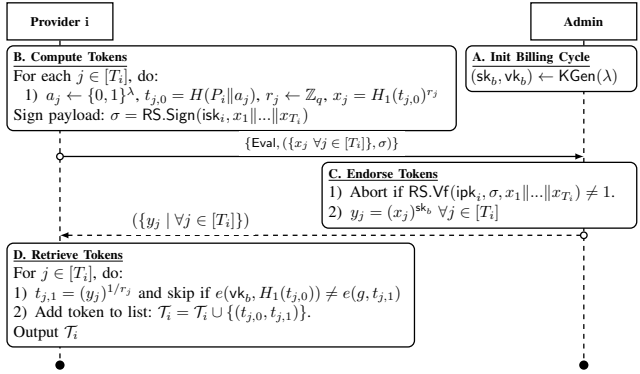

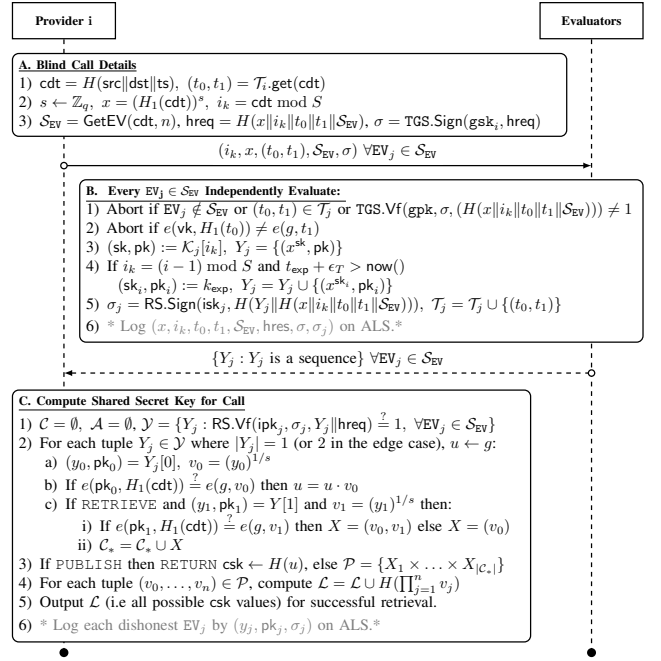
\begin{figure}[t]
    \centering
    \resizebox{.99\columnwidth}{!}{\newlength\CidGenProtoBoxSpacing
\setlength\CidGenProtoBoxSpacing{0.3cm}
\newlength\CidGenProtoLineHeight
\setlength\CidGenProtoLineHeight{-13.6cm}

{
\begin{tikzpicture}[node distance=1cm and 2cm]
\node[actor, anchor=north west, text width=2cm] (Pi) at (0,0) {\textbf{Provider} $\mathbf{i}$};
  \draw[dashed,thick] (Pi.south) -- ++(0,\CidGenProtoLineHeight) node[lifeline] {};
  \node[protobox, anchor=north west, text width=12cm] (PiB1) at ($(Pi.south west)+(0,-\CidGenProtoBoxSpacing)$) {
    \protoboxtitle{A.~Blind Call Details}
    \begin{enumerate}[itemsep=0pt,topsep=0pt,leftmargin=*]
        \item $\calldt = \hash(\callstr),~(t_0, t_1) = \tks_i.\mathsf{get(\calldt)}$
        \item $s \leftarrow \ZZ_q, ~x = (H_1(\calldt))^s, ~\ik = \calldt \bmod\keylistsize$
        \item $\SSS_\mev = \findevals(\calldt, \evparam)$, $\provReqPyld = \hash(\x \| \ik \| t_0 \| t_1 \| \SSS_\mev)$, $\sigma = \gsign{\gsk_i}{\provReqPyld}$
    \end{enumerate}
  };
  \node[protobox, text width=11.2cm, anchor=north west] (PiB2) at ($(PiB1.south west)+(0,-5.6cm)$) {
    \protoboxtitle{C.~Compute Shared Secret Key for Call}
    \begin{enumerate}[itemsep=0pt,topsep=0pt,leftmargin=*]
        \item $\cskset = \emptyset,~\mathcal{A}=\emptyset$, $\filteredfxset = \{\fxset_j : \rsverify{\ipk_j}{\sigma_j}{\fxset_j\|\provReqPyld} \equal 1,~ \forall\mev_j \in \SSS_\mev\}$
        \item For each tuple $\fxset_j \in \filteredfxset$ where $|\fxset_j|=1$ (or $2$ in the edge case), $u\gets g$:
            \begin{enumerate}[itemsep=0pt,topsep=0pt,leftmargin=*]
                \item $(\fx_0, \pk_0) = \fxset_j[0],~v_0 = (\fx_0)^{1/s}$
                \item If $e(\pk_0, H_1(\calldt)) \equal e(g, v_0)$ then $u = u \cdot v_0$ 
                \item If \texttt{RETRIEVE} and $(\fx_1, \pk_1) = \fxset[1]$ and $v_1 = (\fx_1)^{1/s}$ then:
                \begin{enumerate}[itemsep=0pt,topsep=0pt,leftmargin=*]
                    \item If $e(\pk_1, H_1(\calldt)) \equal e(g, v_1)$ then $X = (v_0, v_1)$ else $X = (v_0)$
                    \item $\cskset_* = \cskset_*\cup X$
                \end{enumerate}
            \end{enumerate}
        \item If \texttt{PUBLISH} then \texttt{RETURN} $\csk \gets H(u)$, else $\mathcal{P} = \{X_1 \times \ldots \times X_{|\cskset_*|}\}$
        \item For each tuple $(v_0,\ldots,v_\evparam) \in \mathcal{P}$, compute $\mathcal{L} = \mathcal{L} \cup H(\prod_{j=1}^{\evparam} v_j)$
        \item Output $\mathcal{L}$ (i.e all possible $\csk$ values) for successful retrieval. 
        \item \sendfeedback[Log each dishonest $\mev_j$ by]{$(\fx_j, \pk_j, \sigma_j)$}
    \end{enumerate}
  };
  
\node[actor, anchor=north east, text width=2cm] (EV) at (14cm,0) {\textbf{Evaluators}};
  \draw[dashed,thick] (EV.south) -- ++(0,\CidGenProtoLineHeight) node[lifeline] {};
  
  \node[protobox, anchor=north east, text width=12.2cm] (EVB1) at ($(EV.south east)+(0,-3.1cm)$) {
    \protoboxtitle{B.~ Every $\mathbf{\mev_j \in \SSS_\mev}$ Independently Evaluate: }
    \begin{enumerate}[itemsep=0pt,topsep=0pt,leftmargin=*]
        \item Abort if $\mev_j \notin \SSS_\mev$ or $(t_0, t_1) \in \tks_j$ or $\gverify{\gpk}{\sigma}{\hash(\x \| \ik \| t_0 \| t_1 \| \SSS_\mev)} \neq 1$
        \item Abort if $e(\vk, H_1(t_0)) \neq e(g, t_1)$
        \item $(\sk, \pk) := \keylist_{j}[\ik],~\fxset_j = \{(x^{\sk}, \pk)\}$
        \item If $\ik = (\expidx-1)\bmod\keylistsize$ and $\texp+\talive>\tnow$\par
        ~~$(\sk_{\expidx}, \pk_{\expidx}):=\expkey,~\fxset_j = \fxset_j \cup \{(x^{\sk_{\expidx}}, \pk_{\expidx})\}$
\item $\sigma_j = \rssign{\isk_j}{H(\fxset_j\|\hash(\x \| \ik \| t_0 \| t_1 \| \SSS_\mev))},~\tks_j = \tks_j \cup \{(t_0, t_1)\}$
        \item \sendfeedback{$(\x, \ik, t_0, t_1, \SSS_\mev, \cpsResPyld, \sigma, \sigma_j)$}
    \end{enumerate}
  };

\draw[->, >=latex, thick, msg req]
    ($(Pi.south)+(0,-2.8cm)$) 
    -- 
    ($(EV.south)+(0,-2.8cm)$) 
    node[pos=0.5, above,fill=white] {
$(\ik, x, (t_0, t_1), \SSS_\mev, \sigma)~\forall \mev_j\in\SSS_\mev$
    };
    
\draw[dashed, <-, >=latex, thick, msg res]
    ($(Pi.south)+(0,-7.4cm)$) 
    -- 
    ($(EV.south)+(0,-7.4cm)$) 
    node[pos=0.5, above,fill=white] {
$\{ \fxset_j : \fxset_j~\text{is a sequence}\}~\forall\mev_j\in\SSS_\mev$
    };
\end{tikzpicture}
}
 }
    \caption{\textbf{\emph{\cidgenproto Protocol}} enables providers to establish a shared, ephemeral secret $\csk$ for each call. $\provider_i$ blinds the call descriptor $\calldt$ with a random exponent and sends it to a subset $(\evparam \ll \numevs)$ of \EVs closest to $\calldt$, authenticated by a group signature $\sigma$. Each \EV checks recipient status, verifies $\sigma$, validates the token, and returns $\fx_j$, while also sending feedback to the \LS. $\provider_i$ computes and verifies $\csk$, and can report misbehaving \EVs to the \LS.}
\label{fig:cidgen}
\end{figure}

\begin{figure}[t]
    \centering
\resizebox{.99\columnwidth}{!}{\newlength\RetrieveProtoBoxSpacing
\setlength\RetrieveProtoBoxSpacing{0.4cm}
\newlength\RetrieveProtoLineHeight
\setlength\RetrieveProtoLineHeight{-10.2cm}

{
\begin{tikzpicture}[node distance=1cm and 2cm]
\node[actor, anchor=north west, text width=2cm] (Pi) at (0,0) {\textbf{Provider} $\mathbf{i}$};
  \draw[dashed,thick] (Pi.south) -- ++(0,\RetrieveProtoLineHeight) node[lifeline] {};
  \node[protobox, anchor=north west, text width=10cm] (PiB1) at ($(Pi.south west)+(0,-1.1cm)$) {
    \protoboxtitle{B.~Authenticated Retrieval.}
    \begin{enumerate}[itemsep=0pt,topsep=0pt,leftmargin=*]
        \item $\idx = \hash(\csk)$, $(t_0, t_1) = \tks_i.\mathsf{get(\calldt)}$, $\SSS_\mms = \findms(\csk, \replparam)$
        \item $\provReqPyld = \hash(\idx) \| \hash(t_0 \| t_1 \| \SSS_\mms)$, $\sigma_i = \gsign{\gsk_i}{\provReqPyld}$
    \end{enumerate}
  };
  
  \node[protobox, anchor=north west, text width=10cm] (PiB2) at ($(PiB1.south west)+(0,-4.7cm)$) {
    \protoboxtitle{D.~Message Decryption.}
    For $\forall j \in [\replparam]$, do:
    \begin{enumerate}[itemsep=0pt,topsep=0pt,leftmargin=*]
        \item Skip if $\rsverify{\ipk_j}{\sigma_r}{\provReqPyld\|H(\idx\|c_0\|c_1\|bb\|\sigma)} \neq 1$
        \item Skip if $\gverify{\gpk}{\sigma}{\hash(\idx \| c_0 \| c_1) \| bb} \neq 1$
        \item $\msg = \sum.\dec(\hash(c_0 \| \csk), c_1)$ and return $\msg$ if valid.
        \item \sendfeedback[Optionally report]{invalid responses}
    \end{enumerate}
  };
  
\node[actor, anchor=north west, text width=2cm] (EV) at (7cm,0) {\textbf{Evaluators}};
\draw[dashed,thick]
  (EV.south) -- ++(0,-0.9cm)
   node[lifeline] {};

\node[actor, anchor=north east, text width=2cm] (MS) at (13.5cm,0) {\textbf{Message Stores}};
  \draw[dashed,thick] (MS.south) -- ++(0,\RetrieveProtoLineHeight) node[lifeline] {};
  
  \node[protobox, anchor=north east, text width=11.8cm] (MSB1) at ($(MS.south east)+(0,-3.5cm)$) {
    \protoboxtitle{C.~Each $\mms_j \in \SSS_\mms$ will do:}
    \begin{enumerate}[itemsep=0pt,topsep=0pt,leftmargin=*]
        \item $\provReqPyld = \hash(\idx) \| \hash(t_0 \| t_1 \| \SSS_\mms)$
        \item Abort if $\mms_j \notin \SSS_\mms$ or $(t_0, t_1) \in \tks_j$ or $\gverify{\gpk}{\sigma_i}{\provReqPyld} \neq 1$
        \item Abort if $e(\vk_b, H_1(t_0)) \neq e(g, t_1)$ or $\idx \notin \mmsdb_j$
        \item $\cpsResPyld = H(\idx\|c_0\|c_1\|bb\|\sigma)$, $\sigma_r = \rssign{\isk_j}{\provReqPyld\|\cpsResPyld}$, $\tks_j = \tks_j \cup (t_0, t_1)$
        \item \sendfeedback{$(H(\idx), t_0, t_1, \SSS_\mms, \cpsResPyld, \sigma_i, \sigma_r)$}
    \end{enumerate}
  };

\draw[->, >=latex, thick, msg req]
    ($(Pi.south)+(0,-0.7cm)$) 
    -- 
    ($(EV.south)+(0,-0.7cm)$) 
    node[pos=0.5, above] {
        \textbf{A.~Run \cidgenproto (Fig.~\ref{fig:cidgen}).}
    };
    
\draw[->, >=latex, thick, msg req]
    ($(Pi.south)+(0,-3.3cm)$) 
    -- 
    ($(MS.south)+(0,-3.3cm)$) 
    node[pos=0.5, above,fill=white] {
$\{\mathbf{\mathsf{Retrieve}}, (\idx, (t_0, t_1), \SSS_\mms, \sigma_i)\}~\forall \mms_j \in \SSS_\mms$
    };
    
\draw[dashed, <-, >=latex, thick, msg res]
    ($(Pi.south)+(0,-7cm)$) 
    -- 
    ($(MS.south)+(0,-7cm)$) 
    node[pos=0.5, above,fill=white] {
$(\idx, (c_0, c_1), bb, \sigma, \sigma_r)~\forall \mms_j \in \SSS_\mms$
    };
\end{tikzpicture}
}
 }
    \caption{\textbf{\emph{\retproto}} lets $\provider_i$ retrieve messages for a specific call. After running \cidgenproto to derive $\csk$ and $\idx$, the provider queries the $\replparam$ \MSs for the record. Each \MS returns the encrypted message if present. $\provider_i$ then verifies responses, discards invalid entries, and decrypts to recover $\msg$.}
\label{fig:retrieve}
\end{figure}
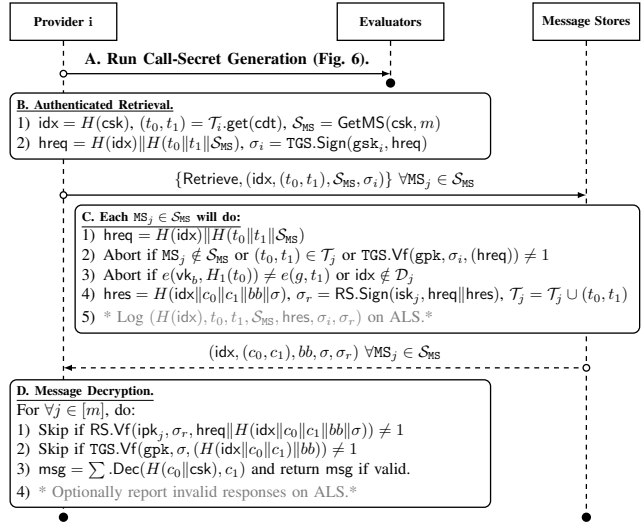

\mysubsubsection{\sysname Protocol Overview}
We present the detailed interaction flow for each \sysname sub-protocol, with comments provided in each caption. Fig.~\ref{fig:protosetup} (\setupproto) shows system initialization; Fig.~\ref{fig:tokens} (\tokensproto) covers token minting; Fig.~\ref{fig:cidgen} (\cidgenproto) details shared ephemeral call secret generation; Fig.~\ref{fig:publish} (\pubproto) illustrates record publishing; and Fig.~\ref{fig:retrieve} describes record retrieval and decryption. Gray text (e.g., {\color{gray}* Log [feedback] to \LS.*}) in the figures indicates that feedback is decoupled from the main call flow and should be deferred to the background to avoid impacting call latency.

\begin{figure}[t]
    \centering
\resizebox{.99\columnwidth}{!}{\newlength\PublishProtoBoxSpacing
\setlength\PublishProtoBoxSpacing{0.4cm}
\newlength\PublishProtoLineHeight
\setlength\PublishProtoLineHeight{-8.2cm}

{\em 
\begin{tikzpicture}[node distance=1cm and 2cm]
\node[actor, anchor=north west, text width=2cm] (Pi) at (0,0) {\textbf{Provider} $\mathbf{i}$};
  \draw[dashed,thick] (Pi.south) -- ++(0,\PublishProtoLineHeight) node[lifeline] {};
  \node[protobox, anchor=north west, text width=10cm] (PiB1) at ($(Pi.south west)+(0,-1.1cm)$) {
    \protoboxtitle{B.~Authenticated Encryption}
    \begin{enumerate}[itemsep=0pt,topsep=0pt,leftmargin=*]
        \item $\idx = \hash(\csk)$, $c_0 \leftarrow \{0, 1\}^\lambda$,  $c_1 = \sum.\enc(\hash(c_0 \| \csk), \msg)$.
        \item $\SSS_\mms = \findms(\csk, \replparam)$, $(t_0, t_1) = \tks_i.\mathsf{get(\calldt)}$ 
        \item $\provReqPyld = \hash(\idx \| c_0 \| c_1) \| H(t_0 \| t_1 \| \SSS_\mms)$, $\sigma = \gsign{\gsk_i}{\provReqPyld}$
    \end{enumerate}
  };

\node[actor, anchor=north west, text width=2cm] (EV) at (6.5cm,0) {\textbf{Evaluators}};
  \draw[dashed,thick] (EV.south) -- ++(0,-0.9cm) node[lifeline] {};

\node[actor, anchor=north east, text width=2cm] (MS) at (13cm,0) {\textbf{Message Stores}};
  \draw[dashed,thick] (MS.south) -- ++(0,\PublishProtoLineHeight) node[lifeline] {};
  
  \node[protobox, anchor=north east, text width=10.5cm] (MSB1) at ($(MS.south east)+(0,-4.0cm)$) {
    \protoboxtitle{C.~Each $\mms_j \in \SSS_\mms$ Caches Record for $\tmax$ seconds.}
    \begin{enumerate}[itemsep=0pt,topsep=0pt,leftmargin=*]
        \item $\provReqPyld = \hash(\idx \| c_0 \| c_1) \| H(t_0 \| t_1 \| \SSS_\mms)$ 
        \item Abort if $\gverify{\gpk}{\sigma}{\provReqPyld} \neq 1$ or $\mms_j \notin \SSS_\mms$ or $(t_0, t_1) \in \tks_j$. 
        \item Abort if $e(\vk_b, H_1(t_0)) \neq e(g, t_1)$
        \item $bb = \hash(t_0 \| t_1 \| \SSS_\mms),~\mmsdb_j[\idx] = ((c_0, c_1), bb, \sigma)$, $\tks_j = \tks_j \cup (t_0, t_1)$
        \item $\sigma_r = \rssign{\isk_j}{\provReqPyld\|\mathsf{ok}}$ and Delete $\mmsdb_j[\idx]$ after $\tmax$ seconds.
        \item \sendfeedback{$(H(\idx\|c_0\|c_1), t_0, t_1, \SSS_\mms, \sigma, \sigma_r)$}
    \end{enumerate}
  };

\draw[->, >=latex, thick, msg req]
    ($(Pi.south)+(0,-0.7cm)$) 
    -- 
    ($(EV.south)+(0,-0.7cm)$) 
    node[pos=0.5, above] {
        \textbf{A.~Run \cidgenproto (Fig.~\ref{fig:cidgen}).}
    };
    
\draw[->, >=latex, thick, msg req]
    ($(Pi.south)+(0,-3.7cm)$) 
    -- 
    ($(MS.south)+(0,-3.7cm)$) 
    node[pos=0.5, above,fill=white] {
       $\{\mathbf{\mathsf{Store}}, (\idx, c_0, c_1, (t_0, t_1), \SSS_\mms, \sigma)\}~\forall \mms_j \in \SSS_\mms$
    };
    
\draw[dashed, <-, >=latex, thick, msg res]
    ($(Pi.south)+(0,-7.9cm)$) 
    -- 
    ($(MS.south)+(0,-7.9cm)$) 
    node[pos=0.5, above,fill=white] {
$\mathbf{(\mathsf{ok}, \sigma_{r})}~\forall \mms_j \in \{\mms_1, ..., \mms_{\replparam}\}$
    };
\end{tikzpicture}
}
 }
    \caption{\textbf{\emph{\pubproto Protocol}} lets $\provider_i$ publish a message $\msg$ to a subset of \MSs. After running \cidgenproto to derive $\csk$, the provider encrypts $\msg$ and sends the ciphertext, token, and group signature $\sigma$ to the $\replparam \ll \nummss$ \MSs closest to $\csk$. Each \MS checks if it is a designated recipient, verifies $\sigma$, validates the token, caches the payload for $\tmax$ seconds if valid, and submits feedback to the \LS.}
\label{fig:publish}
\end{figure}
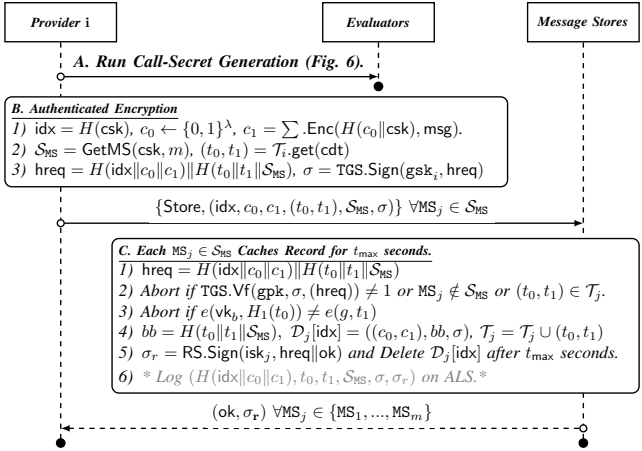

\subsection{\sysname Security Guarantees}
In this section, we argue informally that \sysname achieves the security properties outlined in Sec.~\ref{sec:requirements}. The formal proof in the UC framework is presented in Appendix~\ref{sec:formal-desc}.

\begin{theorem}\label{thm:main-informal}
    Assuming the security of group signature schemes, the security of Oblivious Pseudorandom Function protocol, unforgeability of the signature schemes, and secure hash functions, \sysname achieves Individual Subscriber Privacy, Call Unlinkability, Trade Secrecy, \SecReqLocationConfidentiality, \SecReqRecordExpiry, Perfect Forward Secrecy, and Post-Compromise Security. 
\end{theorem}

\myparagraphi{Individual Subscriber Privacy}
All keys and the $\csk$ used to encrypt and store a record are derived deterministically from the call details themselves. Consequently, records remain protected except in the unlikely event that an adversary can guess the exact call details within $\tmax$ seconds.

\myparagraphi{Call Unlinkability}
The details linking a caller to a recipient are never revealed in plaintext. The only publicly visible values are ciphertexts $(c_0, c_1)$, an index $\idx$, and the set of nodes $\mms_j$. All of these are derived from $\csk$ which is computed via an OPRF protocol. By the obliviousness property of the OPRF scheme, the underlying call details remain hidden.

\myparagraphi{Trade Secrecy}
Group signatures hide which providers contribute records, which themselves are encrypted, hiding all information about the corresponding calls. Decryption is only possible if an adversary can guess the precise call details for calls in progress. For past calls that have terminated, the records cannot be decrypted if at least one $\mev$ is honest. This is so because each honest $\mev$ updates its keys at regular intervals. 

\myparagraphi{\SecReqLocationConfidentiality}
The storage location of an encrypted record is determined by a function that takes the $\csk$ as input. Since the $\csk$ is computable only by those who know the corresponding call details, the storage location remains hidden from others.

\myparagraphi{\SecReqRecordExpiry}
An honest message store $\mms$ deletes each stored record after a fixed expiry duration. Furthermore, each $\mev$ periodically rotates its keys. Even if a message store is compromised and retains all encrypted records, as long as at least one $\mev$ is honest and erases its old keys, the adversary cannot recover the keys or $\csk$ values required to decrypt the records beyond their expiration.

\myparagraphi{Perfect Forward Secrecy}
Since honest parties regularly erase their secret state, even if they are later compromised, the adversary learns nothing about encrypted data protected by erased keys.

\myparagraphi{Post-Compromise Security}
As noted above, $\mev$ nodes periodically rotate their keys. Thus, if an adversary compromises an $\mev$ at time $t$ and learns the keys used in that interval, it can use the compromised keys to decrypt ciphertexts generated only in that interval, and not ciphertexts generated after $t$.

\section{Implementation}
We implemented the cryptographic primitives as an open-source C++ library \thislib, with Python bindings, and used it in our prototype. 

\myparagraphi{Cryptographic Primitives}  
We implemented the OPRF protocol using \texttt{mcl} over a BN elliptic curve. We used \texttt{XChaCha20} from \texttt{libsodium} for symmetric encryption, IBM's \texttt{libgroupsig} for the \texttt{BBS04} group signature scheme~\cite{boneh2004short}, and the \texttt{x509} module in the \texttt{cryptography} library for \fstirsha and \oobss.

\myparagraphi{\sysname's Prototype}  
We implemented \EVs and \MSs as containerized \texttt{FastAPI}~\cite{fastapi} servers with key rotation and audit logging. Providers parallelize HTTP requests to $\evparam \ll \numevs$ \EVs and $\replparam \ll \nummss$ \MSs, treating each as a single logical operation.

\myparagraphi{\oobss's Prototype}  
Our \oobss prototype implements Certificate Repositories (\sticr{s}) and Call Placement Services (\sticps{s}) as containerized \texttt{FastAPI} servers. To reflect real-world behavior, \sticps{} cache certificates, use keep-alive sessions, and parallelize inter-\sticps{} HTTP requests.

\myparagraphi{Data Generation}\label{sec:tel-network-gen}
Because real telephone network topologies are unavailable, we generated a scale-free network using Jäger’s model~\cite{adei2024jaeger}, capturing structure via preferential attachment, market fitness, and inter-carrier agreements, and extended it for current \fstirsha adoption.

We estimated \stirsha deployment using the Robocall Mitigation Database (RMD)~\cite{robomitidb}, which, as of January 21, 2025, listed 7,346 U.S. providers, with 55.96\% having deployed \stirsha. Since larger providers adopt earlier, we sampled 55.96\% of nodes with probabilities proportional to node degree, reflecting realistic adoption. We then computed all-pairs shortest paths to enumerate call routes.

\begin{figure*}[t!]
  \begin{minipage}[t]{0.99\textwidth}
    \centering
    \captionsetup[subfigure]{justification=centering}
    \subfloat[\sysname delivers end-to-end latency comparable to the non-secure \oobss, despite extensive cryptography.]
    {\includegraphics[width=0.30\linewidth]{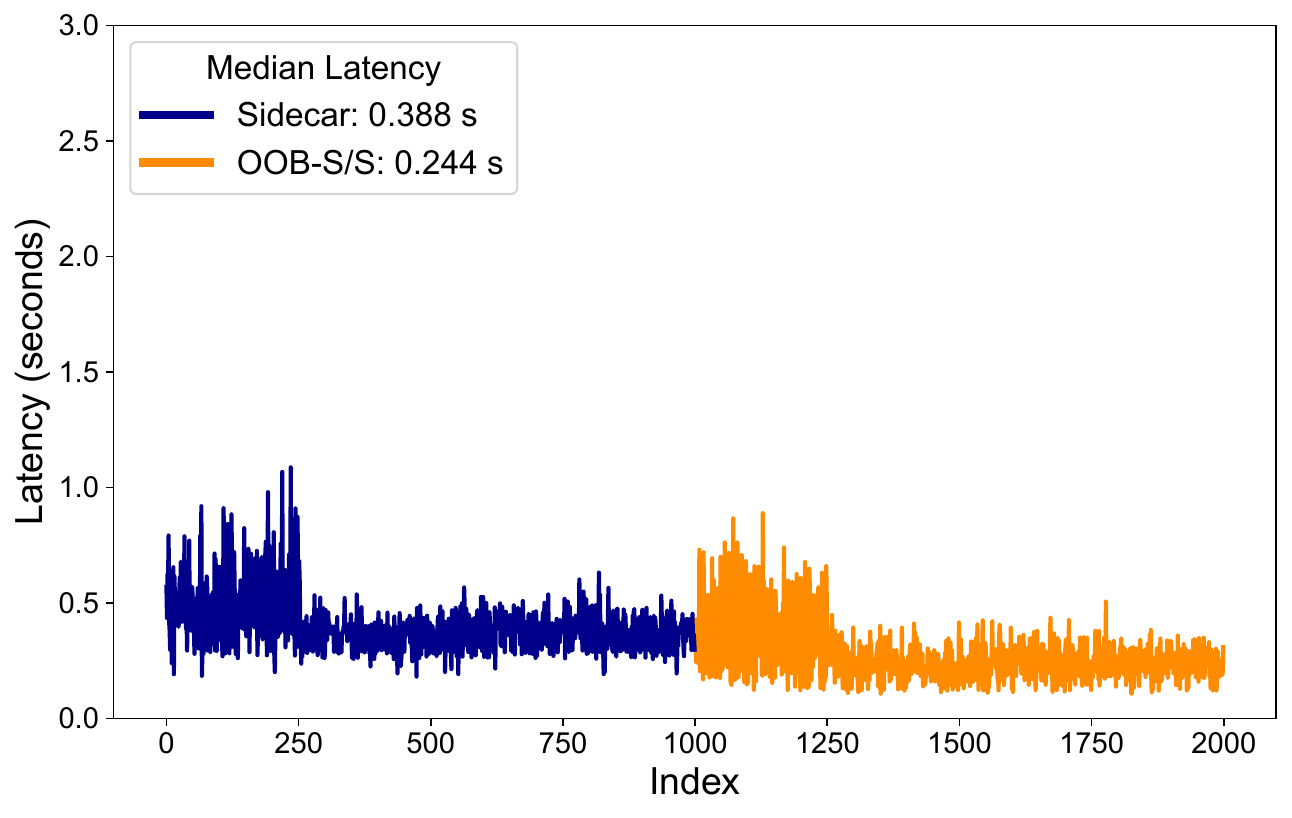}
    \label{fig:latency-trends}}
    \hfill
    \subfloat[\sysname sustains lower response times as call volume increases, demonstrating superior scalability.]
    {\includegraphics[width=0.36\linewidth]{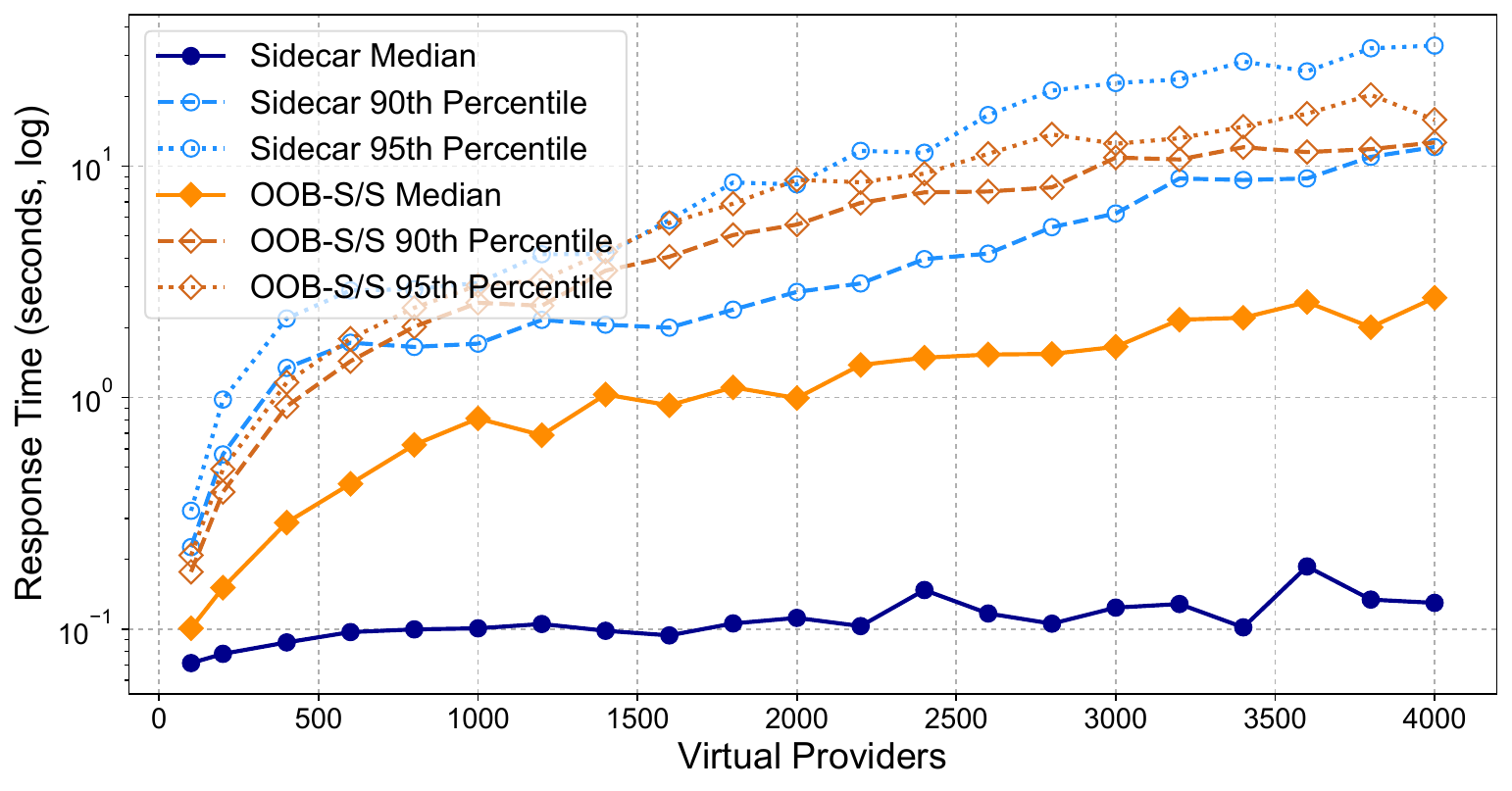}
    \label{fig:res-time}}
    \hfill
    \subfloat[\sysname maintains higher throughput and degrades more gracefully than \oobss as call rates increase.]
    {\includegraphics[width=0.307\linewidth]{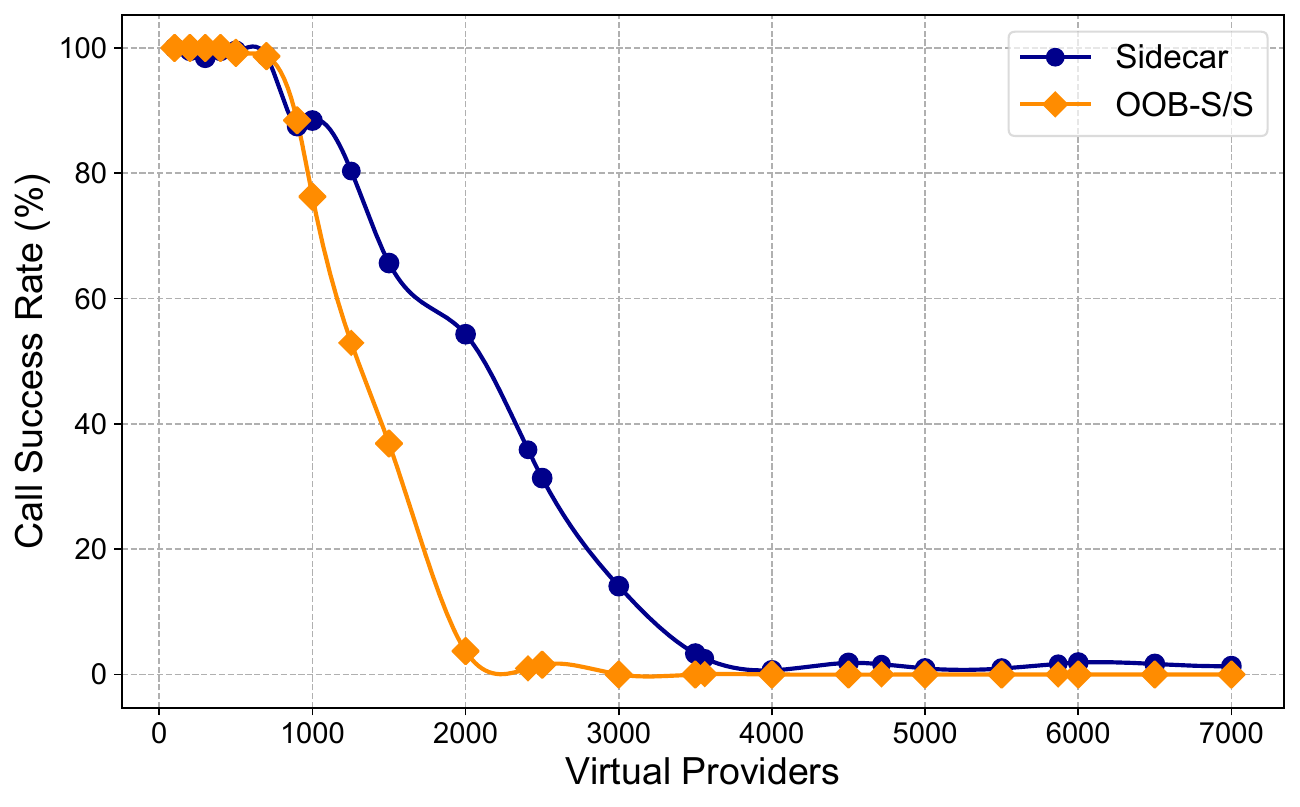}
    \label{fig:throughput}}

\caption{System performance results comparing \sysname and \oobss}
    \label{evaluation-charts}
  \end{minipage}
\end{figure*}

\myparagraphi{\sysname Inter-Work Function (\iwf)}  
We developed \iwf, a lightweight gateway plugin for seamless \sysname deployment in telephony networks. To evaluate real-world integration challenges, we built a physical testbed with four provider nodes interconnected via SIP and TDM trunks (see Appendix~\ref{sec:iwf}).

 \section{Evaluation}
We evaluate \sysname in three experiments. Experiment~1 identifies $(\evparam, \replparam)$ pairs that balance security and resilience. Experiment~2 estimates resource requirements—vCPUs, memory, bandwidth, and storage—for \sysname \cpss. Experiment~3 compares \sysname's scalability and latency with \oobss.

\subsection{Results}
We first present our key takeaways, then detail each experiment in the following sections.

\takeaway{Despite the extensive cryptographic guarantees, \sysname adds only a fraction of a second to the latency experienced by subscribers placing and receiving calls.} 
Fig.~\ref{fig:latency-trends} shows the end-to-end latency added by \sysname, reflecting total computational and communication cost across all provider hops. Most calls incur less than 1 second of extra delay, with a median of about 0.4 seconds—roughly equivalent to a blink of an eye and imperceptible to users. Since a typical call setup already takes several seconds, this added latency is negligible.

\takeaway{Given the same resources, \sysname provides significantly better response times and higher throughput compared to \oobss.} 
Fig.~\ref{fig:res-time} shows that as call volume increases, \sysname maintains a low median response time and consistently outperforms \oobss, whose latency rises with load. \sysname's tail latencies (90th and 95th percentiles) are also lower or comparable to \oobss. Fig.~\ref{fig:throughput} shows delivery success rates under load: \sysname degrades gracefully, maintaining 54.31\% success at 2,000 virtual providers, while \oobss drops sharply to 3.74\% due to latency spikes exceeding the 3-second timeout.

\takeaway{\sysname gives ``six nines'' uptime on commodity cloud infrastructure.} 
Using AWS EC2 compute instances with a minimum 99.0\% guaranteed availability, \sysname offers 99.9999\% availability --- just 86.4 milliseconds daily downtime.

\takeaway{\sysname requires only modest compute and bandwidth resources --- just \$25 per an Evaluator or a Message Store, and \$35 for a median provider --- to support 2 billion daily calls across the U.S.}
Table~\ref{table:resource-reqs} summarizes estimated resource requirements per \sticps role. An \EV{} needs 11~vCPUs, 7~GB RAM, 30~Mbps bandwidth; an \MS{} needs 10~vCPUs, 7~GB RAM, 100~Mbps. A provider handling 1,000 calls/sec requires 29~vCPUs, 23~GB RAM, 360~Mbps. Both \EV{} and \MS{} need 71~GB storage for tokens. Costs are based on AWS EC2 On-Demand pricing (US East, Ohio).

\begin{figure}[t]
    \centering
    \includegraphics[width=.80\columnwidth]{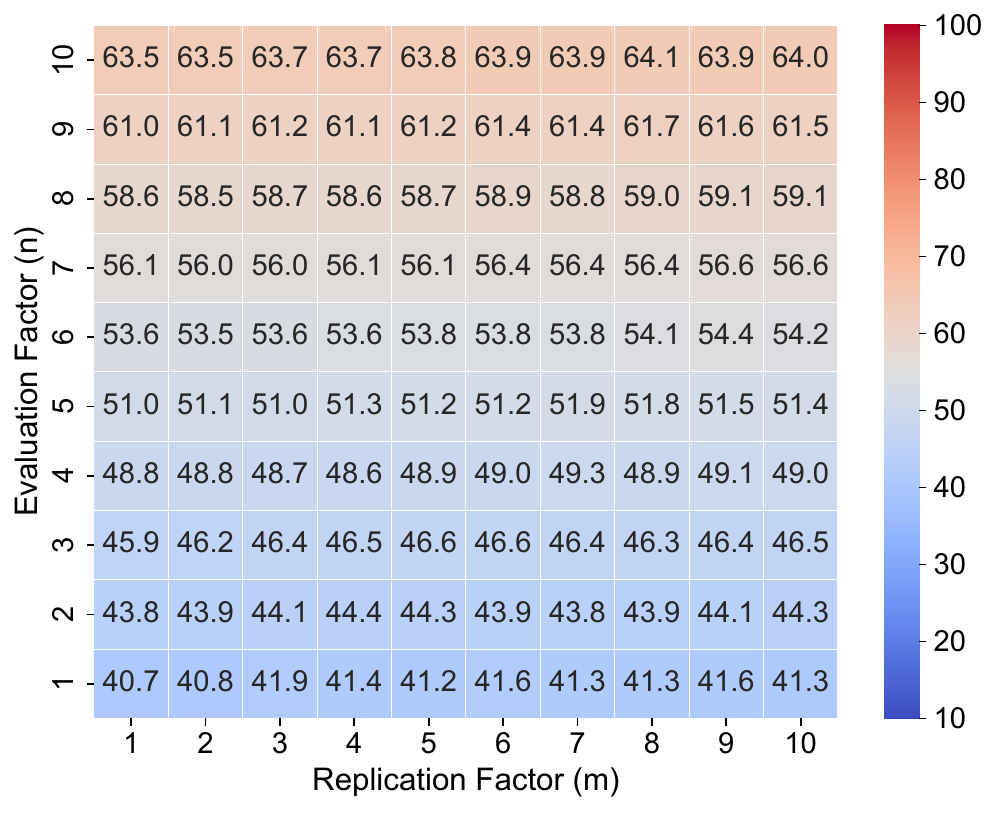}
    \caption{The cryptographic overhead (in milliseconds), which \sysname adds to call setup, is invariant to scaling $(\evparam, \replparam)$}
	\label{fig:latency-heatmap}
\end{figure}

\takeaway{Adding more nodes to \sysname improves both performance and security. In contrast, \oobss's performance degrades at scale.} As shown in Fig.\ref{fig:scaling}, because \oobss uses broadcast---one publish, \(Q-1\) republish, and one retrieve, where $Q$ is the number of \sticps{s}---it forces each node to process nearly one request per call as the number of nodes \(Q\) grows. In contrast, each call in \sysname only requires \(2(\evparam+\replparam)/Q\) requests per node (\(12/Q\) for \(\evparam=3\) and \(\replparam=3\)), so the per-node burden diminishes with increasing $Q$. \sysname's uniform load distribution accelerates performance at larger \(Q\), and it also raises the cost of compromising a threshold of nodes, thereby enhancing security. 

\begin{center}
    \resizebox{.99\columnwidth}{!}{{
\begin{tikzpicture}
\begin{axis}[
    width=13cm, height=4cm,
    domain=1:16, samples=100,
    xlabel={Number of \cps Operators}, ylabel={Per \cps Burden},
    label style={font=\large},
    tick label style={font=\large},
    legend style={at={(0.99,0.7)}, anchor=north east, draw=gray, fill=white},
    grid=both,
    ymin=0, ymax=12,
    xmin=1, xmax=16,
]
    \addplot[very thick, color=blue!70!black] {12 / x};
    \addlegendentry{\sysname}
    \addplot[very thick, color=orange!100!black] {x - 1};
    \addlegendentry{\oobss}
\end{axis}
\end{tikzpicture}
}
     }
    \captionof{figure}{\sysname Outperforms \oobss at Scale}
    \label{fig:scaling}
\end{center}

\subsection{Evaluation Methodology}
We describe our experimental setup, design, and evaluation.

\mysubsubsection{Experimental Setup}
Our control node is a Linux VM with 32~vCPUs and 62~GB RAM, hosted on a Supermicro server (Intel Xeon Gold 6130, ECC DDR memory, 12~Gbps SAS drive). For Experiment~3, we used 10 general-purpose AWS EC2 t3.small instances (2~vCPUs, 2~GB RAM, EBS-only storage, up to 5~Gbps network) across Northern Virginia, Ohio, Oregon, and Northern California. We provisioned infrastructure with Terraform and automated the deployment of \EV{s}, \MS{s}, \sticr{s}, and \sticps{s} using Ansible~\cite{ansible}.

\mysubsubsection{System Resiliency}
This experiment evaluates which $(\evparam, \replparam)$ configurations offer sufficient resilience by balancing trust distribution, availability, and latency. We say $(\evparam, \replparam)$ is ``good enough'' if it meets subjective thresholds for latency and security, as determined by the implementor's threat model.

We used our network model generator to simulate $\sim$1,000 unique calls for each $(\evparam, \replparam)$ pair, with $\evparam, \replparam \in [1, 10]$, recording median end-to-end compute latencies. This experiment primarily measures the cryptographic overhead per call.

Fig.~\ref{fig:latency-heatmap} shows median latency (ms) for various $(\evparam, \replparam)$ configurations. The cryptographic overhead is independent of $\replparam$, with latency remaining nearly constant as $\replparam$ increases. In contrast, latency rises by about 2~ms per additional $\evparam$, due to the higher computational cost of \cidgenproto compared to \pubproto and \retproto.

We select $(\evparam=3, \replparam=3)$ as ``good enough'' for balancing latency and reliability. With independent node failures and 99.00\% availability per AWS instance, the probability all three are down is $(1-0.99)^3 = 10^{-6}$, yielding 99.9999\% system availability. While this choice depends on operational needs, our analysis demonstrates it is both effective and practical.

\begin{table}[t]
    \centering
    \footnotesize
    \caption{\sysname requires minimal computation, with latencies in the millisecond range for $(\evparam=3,\replparam=3)$.}
    \label{table:compute-times}
    \begin{tabular}{|l|r|r|r|r|}
        \hline
        \textbf{Operation} & \textbf{Min} & \textbf{Max} & \textbf{Median} & \textbf{MAD} \\ \hline
        \multicolumn{5}{|l|}{\textcolor{gray}{\cidgenproto}} \\\hline
        Evaluator & 4.500 &  9.447  & 5.267 & 0.140 \\ \hline
        \multicolumn{5}{|l|}{\textcolor{gray}{\pubproto}} \\\hline
        Provider & 10.013 & 23.947 & 13.052 & 0.923 \\
        Message Store & 4.077 & 10.707 & 4.976 & 0.057 \\\hline
        \multicolumn{5}{|l|}{\textcolor{gray}{\retproto}} \\\hline
        Provider & 13.730 & 31.628 & 17.230 & 1.076 \\
        Message Store & 4.059 & 11.368  & 5.006 & 0.059 \\\hline
    \end{tabular}
\normalsize
\end{table}

\begin{table}[t]
    \centering
    \footnotesize
    \caption{\sysname requires modest computing resources.}
    \label{table:resource-reqs}
    \begin{tabular}{|l|r|r|r|r|r|}
        \hline
        \textbf{Entity} & \textbf{vCPUs} & \textbf{Memory} & \textbf{Storage} & \textbf{Bandwidth} \\ \hline
        Evaluator & 11 & 7~GB & 71~GB & 30~Mbps \\ \hline
        Message Store & 10 & 7~GB & 71~GB & 100~Mbps \\ \hline
        Provider & 29 & 23~GB & x & 360~Mbps \\\hline
\end{tabular}
    \normalsize
\end{table}

\mysubsubsection{Resource Requirements} 
We estimate minimum vCPU, memory, storage, and bandwidth requirements for \EV{s}, \MS{s}, and Providers. Lacking ground truth for U.S. multi-carrier call volume, we use available reports to estimate 2 billion such calls daily. Our graph model finds 78\% involve at least one out-of-band hop, so we compute 1.56 billion calls/day for resource estimation. With uniform load distribution, each \EV{} and \MS{} handles $1/\numevs$ and $1/\nummss$ of the total, respectively. We assume providers process calls at a median rate of 1,000 per second.

First, we benchmark $(\evparam=3, \replparam=3)$ with 1,000 iterations to measure compute times for providers, \MS{s}, and \EV{s} across protocols. Second, we deploy a single \EV (4 workers), \MS (4 workers), and \iwf (6 workers) in Docker, and use Grafana k6 to simulate traffic from 1,000 virtual providers over 10 minutes, monitoring memory usage throughout.

Table~\ref{table:resource-reqs} summarizes resource requirements estimated by the following equations: 
\begin{itemize}[itemsep=0pt,topsep=0pt,leftmargin=*]
    \item $n(\textit{vCPU}) = \left\lceil \oobrate \times (\textit{median} + 3\times\textit{MAD}) \right\rceil$
    \item $n(\textit{Memory}) = \left\lceil \frac{\textit{Usage}}{\textit{n(Workers)}} \times O_v \times (2 \cdot \textit{vCPUs} + 1) \right\rceil$
    \item $n(\textit{Storage}) = \left\lceil \frac{\oobrate}{C} \times \tmax \times \textit{AvgRecSize} \times O_v \right\rceil$ 
    \item $n(\textit{Bandwidth}) = \left\lceil \textit{Rate} \times (\textit{Req} + \textit{Res}) \times O_v \right\rceil$
\end{itemize}
We provide the formulas and parameters for estimating the resources for each type in Appendix~\ref{sec:resource-estimation}.

\mysubsubsection{Scalability and End-to-End Latency Overhead} 
\sysname provides stronger security than \oobss but incurs extra cryptographic overhead. Scalability is a key differentiator.

We deployed \sysname and \oobss separately on 10 AWS EC2 instances across multiple U.S. data centers. For \sysname, each instance ran an \EV and an \MS; for \oobss, each hosted a \sticps and its certificate repository (CR). We implemented certificate caching at \sticps{}, enabled keep-alive sessions between \cpss, and retrieved \passports from three \cpss in parallel to improve \oobss's success rates. \cpss use Redis for caching.

We structured the evaluation into three parts. Parts 1 and 2 measured per-request performance using Grafana k6 to send valid, precomputed payloads and collect metrics. In Part~1, we varied the number of clients (providers) from 100 to 4,000 to determine maximum throughput on low-end EC2 instances. Part~2 extended to 6,000 clients with a strict 3-second per-request timeout. Each experiment ran for one minute.

Part~3 measured end-to-end latency. We simulated $\sim$1,000 unique call scenarios using our network model and measured their concurrent execution. Since the requests providers send to individual \EVs and \MSs are independent, we executed them in parallel. Specifically, \sysname performs $\evparam$ (e.g $\evparam=3$) requests in parallel to \EV{s} and $\replparam$ (e.g $\replparam=3$) to \MS{s}, incurring a latency equivalent to only two sequential HTTP rounds. The same applies to \oobss: a provider sends a single request to one \sticps, which in turn issues $Q{-}1$ parallel requests to the others where $Q$ is the number of \sticps{s}, thus completing within two sequential HTTP rounds.

 \section{Discussions}\label{sec:discussion}
We discuss potential applications of \sysname, deployment incentives, and concerns for key stakeholders.

\mysubsubsection{Broader Applications}
\sysname generalizes to a distributed key-value store with cryptographically enforced record expiry. Keys are opaque and derived from live calls; values, currently encrypted \passports, can be any data. Senders do not know recipients, but only intended parties can access records; recipients do not know the sender, but are assured the record pertains to their call. This abstraction extends \sysname's utility beyond \stirsha. We highlight three examples below.

\myparagraphi{Strong Caller Authentication}
While \stirsha enables providers to attest caller authorization for a source number, it does not authenticate the caller’s identity. Emerging systems require end-to-end delivery of authentication metadata. \sysname supports these by enabling reliable, privacy-preserving delivery. For example, Authenticall~\cite{reaves2017authenticall} uses a centralized server for end-to-end authentication and integrity, while UCBlocker~\cite{du2023ucblocker} uses anonymous credentials and recipient-defined blocking, both requiring transmission of authentication data.

\myparagraphi{Collaborative Spam Mitigation}
Providers typically use siloed defenses to block illegal calls, but attackers evade detection by rotating entry points or distributing activity across providers. Effective mitigation may require cross-provider collaboration, which raises privacy concerns~\cite{hu2023collaborative}. \sysname enables privacy-preserving per-call metadata sharing, such as fraud indicators and suspicious patterns. Existing frameworks~\cite{AZAD2019841, azad2017privy, ruan2019cooperative} can integrate with \sysname to securely exchange threat intelligence in real-time and combat evolving telephony abuse.

\myparagraphi{Automated Call Traceback}
Law enforcement faces challenges in tracing the true source of illegal calls. While \stirsha signatures can reveal origins, malicious providers likely will not attest to such calls. Adei et al. proposed a distributed system \cite{adei2024jaeger} for automated traceback regardless of whether the originating provider attested. Integrating with \sysname improves its security, lowers bandwidth and compute costs, and cryptographically ensures that at least one on-path provider must authorize tracebacks—a property the prior protocol only weakly achieves.

\mysubsubsection{Deployment Incentives}
We outline stakeholder incentives.

\myparagraphi{Telecom Operators and Subscribers}
\sysname helps both subscribers and providers by reducing spam, which in turn improves security and restores trust in answering unknown calls. For providers, this translates to fewer revenue-generating calls going unanswered and lower regulatory compliance costs. \sysname also serves as a platform for innovation, enabling any coalition of providers to deploy dedicated instances and offer revenue-generating services like Branded Calling or RCD.

\myparagraphi{RLEAs, IETF STIR WG, and ATIS}
Regulatory and Law Enforcement Agencies (RLEAs), along with standardization bodies like the IETF STIR WG and ATIS, have prioritized mitigating telephony abuse through legal enforcement and protocol development. \sysname aligns with their goals, offering an incrementally deployable tool for future security protocols that require end-to-end metadata delivery.

\myparagraphi{Call Placement Services}
\cps operators earn fees per processed request, incentivizing competitive participation. \sysname's uniform node selection ensures fair traffic distribution and balanced rewards for all \sticps operators.

\mysubsubsection{Deployment Concerns}
Despite \sysname's security and privacy benefits, a successful transition from \oobss requires addressing practical deployment challenges. \sysname's design reflects these concerns. We now discuss concerns for deployments, interoperability, billing, and provider adoption.

\myparagraphi{Deployment Artifacts}
We released Docker images and deployment automation scripts to streamline adoption for all stakeholders. For providers, we developed the \sysname Inter-Work Function (SIWF in Appendix~\ref{sec:iwf}) plugin for easy integration with existing gateways and validated it by integrating SIWF into the Asterisk PBX platform. We believe these artifacts will accelerate \sysname adoption and transition from \oobss.

\myparagraphi{Transitioning and Interoperability with \oobss}
The early state of the \oobss ecosystem, with few large-scale deployments, presents an opportunity for a smooth transition to \sysname. We propose a phased strategy that partitions existing \cps nodes into three groups: a dedicated \sysname group, a legacy \oobss group, and a gateway group. The gateway nodes bridge the two systems by exposing an \oobss-compatible API to nodes in \oobss group while handling all necessary protocol translations, ensuring service continuity as the \oobss group migrates over time.

To explain its function, we first consider a gateway group with a single node. When a provider sends a publish request to a node in \oobss, the gateway—being a peer in \oobss group—receives the request via republish, and runs \sysname's \pubproto to save the \passport in \sysname. For retrievals, if an \oobss node does not have a record, it queries the gateway. The gateway then runs \sysname's \retproto to find and return the record. This ensures that providers can continue using \oobss nodes until they adopt \sysname.

A single-node gateway, however, creates a single point of failure. A resilient bridge therefore requires multiple gateway nodes, and implementing this using standard architectural patterns like load balancing or a leader-elected cluster is a well-understood engineering practice.

\myparagraphi{Recommended Baseline Configuration}
\sysname's architecture can be tuned along a spectrum of decentralization to fit an implementor's threat model. A fully centralized model, where one party manages all subsystems, requires absolute trust in that single entity. A siloed model, where a different dedicated party operates each subsystem, merely creates three distinct single points of security failure. We recommend a baseline configuration of $2$-of-$2$ for both Evaluators and administration, and $2$-of-$\nummss$ for the $\nummss$ available Message Stores. This eliminates single points of failure in each subsystem and provides a robust foundation that can be further decentralized.

\myparagraphi{No New Provider Requirements}
To minimize adoption friction, \sysname adheres to the established \oobss workflow: providers (including intermediaries) upload \passports before SIP-to-TDM conversion or retrieve them after TDM-to-SIP conversion. While this approach can increase latency on multi-TDM paths, an alternative---requiring only originating and terminating providers to interact with \sysname---would reduce latency but impose the disruptive requirement of mandatory adoption on customer-facing providers. Because \sysname is designed to support both workflows, it can default to the current workflow, thereby placing no new obligations on providers that have already implemented in-band \stirsha or \oobss.

\myparagraphi{Support for Pay-per-Use Billing}
\sysname enables a pay-per-use billing model, a more equitable approach than fixed fees. Usage is tracked via a cryptographic token audit trail. This mechanism is intended not as a replacement but as an auxiliary feature that can integrate with existing billing systems, allowing the industry to adopt this usage-based model. \section{Conclusion}
\label{sec:conclusion}
We introduced \sysname, a privacy-preserving system for secure call metadata transmission across all telephone networks. \sysname is efficient, requiring modest resources and adding minimal call setup latency. It offers a practical, scalable solution for retrofitting fragmented telephony infrastructure with strong privacy, accountability, and tunable decentralization.

\bibliographystyle{abbrv}

\appendix
\subsection{Justification of Requirements for Out-of-Band Signaling}\label{sec:justification}
In this section, we provide justifications for the functional and security requirements we mandate for \emph{out-of-band signaling}.

\mysubsubsection{Functional Requirements}
These functional requirements, while foundational, are non-negotiable in the context of telephony, which operates as real-time, high-availability critical infrastructure. At a minimum, the system must be both useful and correct; it must reliably perform its core function of record upload and lookup (\ref{funcreq:f1}) with correctness (\ref{funcreq:f2}), as incorrect data would undermine the security attestations it is meant to support. Furthermore, it must meet stringent performance criteria. Any new component must be highly efficient (\ref{funcreq:f3}) to avoid adding perceptible latency to call setup, and scalable (\ref{funcreq:f4}) to handle the peak call volumes of the global network without degradation. Finally, as a component of critical infrastructure, the system must be resilient (\ref{funcreq:f5}), ensuring high availability and fault tolerance even in the face of network or node failures. Satisfying all these requirements simultaneously is the central challenge in designing a practical and adoptable out-of-band signaling system.

\mysubsubsection{Security Requirements}
The security requirements are derived from a threat model that considers telephony abuse, mass surveillance, and the business realities of a competitive provider ecosystem. The foundational principle is that no off-path entity should learn anything of substance about a call. The following paragraphs provide a detailed justification for each requirement and explain how it contributes to this core principle.

\myparagraphi{Individual Subscriber Privacy}
This is the most fundamental privacy guarantee. Without it, an adversary could conduct ``fishing expeditions'' by querying the system with a known phone number to see if that individual has made any calls, or to whom. It prevents the system from being used as a directory to confirm the existence of a communication relationship without possessing full knowledge of it beforehand.

\myparagraphi{Call Unlinkability}
Protecting individual records is not sufficient if an attacker can link them together to map out communication patterns. An adversary who compromises a single point in the network should not be able to learn a target's entire social graph by observing their call activity. This requirement directly counters traffic analysis and is critical for protecting sensitive communications from any off-path observer.

\myparagraphi{Trade Secrecy}
For any multi-provider system to be viable, it must protect the business interests of its participants. A provider's call volumes, routing agreements, and customer relationships are highly sensitive trade secrets that can be inferred from traffic analysis. A system that exposes this data would create a strong disincentive for participation. This requirement ensures that the system cannot be used for corporate espionage.

\myparagraphi{Location Confidentiality}
If an off-path adversary can determine which network node stores a specific record, that node becomes a target for a focused denial-of-service attack or a targeted compromise. This requirement ensures that the location of a record is itself a secret, known only to the on-path participants. Hiding the storage location forces an adversary to attack the entire network rather than a single, known target.

\myparagraphi{Record Expiry}
The principle of data minimization \cite{nist-privacy-framework} dictates that sensitive information should not be stored indefinitely. A historical archive of call metadata is a liability that grows over time. This requirement limits the temporal window of any potential data breach; even if the entire system is compromised in the future, access to past records is impossible. While the conventional \oobss standard suggests operators delete records, it provides no cryptographic mechanism to enforce this, leaving data vulnerable to indefinite retention.

\myparagraphi{Forward Secrecy and Post-Compromise Security}
The guarantees of Forward Secrecy (PFS) and Post-Compromise Security (PCS) are standard requirements for modern secure messaging systems and apply equally to out-of-band signaling---ad-hoc message exchanges. PFS ensures that the compromise of long-term keys does not compromise past records, while PCS ensures such a compromise does not permanently break the security of future records. Together, they are essential for containing the temporal impact of a security breach. \subsection{Content Addressing}\label{sec:content-addr}
\sysname{} consists of $\numevs$ \EV{}s and $\nummss$ \MS{}s. Providers interact with both \EV{}s and \MS{}s, but \EV{}s and \MS{}s do not directly interact with each other. Unlike many real-world P2P systems, \sysname{} sees low node churn: only authorized nodes may join, and departures are rare.

The \SOsLong{} maintains multiple replicated copies of a Public Registry, $\routetable$, which holds $O(\numevs + \nummss)$ entries. Each entry is a tuple $(\nodeid, \ipaddr, \nodetype)$, where $\nodeid$ is a unique 256-bit hash, $\ipaddr$ is the node’s IP address, and $\nodetype \in \{\mev, \mms\}$. When a legitimate \EV{} or \MS{} registers, the \SOsLong computes $\nodeid \leftarrow \hash(\ipaddr \|\nodetype \| \{0,1\}^\lambda)$ and add the new record to $\routetable$. To revoke a node, they simply remove the corresponding record from $\routetable$.

Because $\routetable$ remains relatively stable, providers store a local copy for offline discovery and periodically synchronize it to stay up to date. As described in Sec.~\ref{sec:base-sys}, the functions $\findms$ and $\findevals$ return \MS{} and \EV{} records, respectively, and both wrap the generic $\findnodes(x, q, \nodetype)$ in Fig.~\ref{fig:findnodes}. Given an integer $q$, $\findnodes$ returns the $q$ nodes closest to $x$, using XOR distance. The algorithm uses a max-heap to track the $q$ nearest nodes, achieving a time complexity of $O(|\mathcal{N}|\cdot \log q)$. An alternative—sorting all distances and selecting the first $q$—requires $O(|\mathcal{N}|\cdot \log |\mathcal{N}|)$ time. The heap-based approach is more efficient in practice since $q \ll |\mathcal{N}|$.

\begin{figure}[h]
    \centering
    \begin{tcolorbox}[standard jigsaw,opacityback=0]
        \begin{flushleft}
            {\footnotesize
            \smallskip\noindent\underline{$\findnodes(x, q, \nodetype)$}
            \begin{enumerate}[itemsep=0pt, topsep=4pt]
                \item Retrieve $\mathcal{N}$, the set of nodes of type $\nodetype$.
                \item Initialize a max-heap $\mathcal{H}_p$ (ordered by distance $d$) of size $q$.
                \item For each node $\mathsf{nd}_j \in \mathcal{N}$:
                \begin{enumerate}[itemsep=0pt, topsep=-2pt]
                    \item Compute $d_j = \hash(x) \xor \hash(\mathsf{nd}_j.\nodeid)~\text{where}~d_j \in \ZZ$.
                    \item If $|\mathcal{H}_p| < q$, push $(\mathsf{nd}_j, d_j)$ into $\mathcal{H}_p$.
                    \item Else if $d_j < \text{max}(\mathcal{H}_p)$, replace the max with $(\mathsf{nd}_j, d_j)$.
                \end{enumerate}
                \item Return the nodes in $\mathcal{H}_p$ as the $q$ closest nodes to $x$.
            \end{enumerate}
            }
        \end{flushleft}
    \end{tcolorbox}
\caption{The $\findnodes$ algorithm uses a max-heap to find the $q$ closes nodes to the given $x$ based on the XOR-metric}
    \label{fig:findnodes}
\end{figure} \subsection{Estimating the Resource Requirements}\label{sec:resource-estimation}
\mysubsubsection{vCPU} We estimate the minimum number of vCPUs using compute time statistics from Table~\ref{table:compute-times}, incorporating both the median and median absolute deviation (MAD) to account for variability: \vspace{-3pt}
\begin{equation}
    n(\textit{vCPU}) = \left\lceil \oobrate \times (\textit{median} + 3\times\textit{MAD}) \right\rceil
\end{equation}
\noindent where $\oobrate$ is the call rate per node type (\EV, \MS, or provider), and the $3\times\textit{MAD}$ term provides overhead.

\mysubsubsection{Memory} We estimate memory requirements using peak usage data from Experiment~2:
\begin{equation}
    n(\textit{Mem}) = \left\lceil \frac{\textit{Usage}}{\textit{n(Workers)}} \times O_v \times (2 \cdot \textit{vCPUs} + 1) \right\rceil
\end{equation}
\noindent where $(2 \cdot \textit{vCPUs} + 1)$ is the recommended number of workers~\cite{gunicorn_rec_workers}, and $O_v$ accounts for 50\% (1.5) overhead. We observed peak usage of 850~MB for an \MS, 785~MB for an \EV, and 1.5~GB for the \iwf used by providers.

\mysubsubsection{Storage} We estimate the storage needed per Message Store or Evaluator:
\begin{equation}
    n(\textit{Storage}) = \left\lceil \frac{\oobrate}{C} \times \tmax \times \textit{AvgRecSize} \times O_v \right\rceil
\end{equation}
\noindent Here, $C=\nummss$ for \MSs or $C=\numevs$ for \EVs and $O_v$ accounts for 50\% (1.5) overhead.  Message stores retain records only for $\tmax = 15~\text{seconds}$.

\mysubsubsection{Bandwidth} We estimate the required bandwidth as:
\begin{equation}
n(\textit{Bandwidth}) = \left\lceil \textit{Rate} \times (\textit{Req} + \textit{Res}) \times O_v \right\rceil
\end{equation}
\noindent where $O_v$ is the assumed per-request overhead (50\%). The request-response size is 1.3~KB for \cidgenproto, 1.5~KiB for \pubproto, and 2.2~KiB for \retproto. \subsection{A formal UC definition for Out-of-Band Signaling}\label{sec:formal-desc}

In this section, we present an ideal functionality for out-of-band signaling $\foos$. The $\foos$ ideal functionality presented in Figure~\ref{fig:foos} captures all the functional and security requirements defined in Section~\ref{sec:requirements}.

\begin{figure}
    \centering
    \begin{tcolorbox}[standard jigsaw, opacityback=0]

    \ifoobfullversion\else\footnotesize\fi
    
    \myparagraph{Participants} 
    The set of providers $\PPP$ is initialized as an empty set. The set $\MMM \subset \PPP$ that includes the set corrupt parties. 
    
    \myparagraph{Data Structures} A table $\DDD$ initialized as $\emptyset$. A list of generated tokens $\LLL_\tkns$ and a list of redeemed tokens $\LLL_\rdmd$. 

    \myparagraph{Predicates} The functionality has two predicates $\checkexpiry$ that takes as input an entry $x$ in $\DDD$ and outputs $\bot$ if the entry has expired, and outputs $x$ otherwise, and  $\valid$ that takes as input communication messages and outputs $\emptyset$ if all messages generated honestly or the identities of the maliciously behaving parties.  

    \myparagraph{Functionality}

    \begin{itemize}
        \item {\bf Register}: Upon receiving $(\register, P_i)$ from a party $P_i$ do $\PPP := \PPP \cup \{P_i\}$ and send $(\register, P_i)$ to $\adv$. 
        \item {\bf Get Token}: Upon receiving $(\gettkn)$ from some $P_i$ send $(\gettkn, P_i)$ to $\adv$ and get back $\tkn$, store $(P_i, \tkn)$ in $\LLL_\tkns$ and send back $(\gettkn, \tkn)$ to $P_i$. 
        \item {\bf \cidgenproto}: Upon receiving $(\genid, \callstr, \tkn)$ from $P_i$,
        \begin{itemize}
            \item If $\tkn \notin \LLL_\tknlst$ or $\tkn \in \LLL_\rdmd$ ignore, if  else add $(P_i,\tkn)$ to $\LLL_\rdmd$. 
            \item Check if an entry $(\cdot, \callstr, \csk, \cdot)$ exist in $\DDD$ else send $(\genid)$ to $\adv$  and get back $\csk$ and send $(\genid, \callstr, \csk)$ to $P_i$. 
            \item Store $(\cdot, P_i, \callstr, \csk, \cdot)$ in $\DDD$
        \end{itemize}    

        \item {\bf \pubproto}: Upon receiving $(\mpub, \csk, \msg, \tkn)$ from $P_i$: 
        \begin{itemize}
            \item If $\tkn \notin \LLL_\tknlst$ or $\tkn \in \LLL_\rdmd$ ignore, if  else add $(P_i,\tkn)$ to $\LLL_\rdmd$. 
            \item Check if there exists an entry in $\DDD$ with $(\cdot, \cdot,\cdot,\csk,\cdot)$. 
            
            \item If an entry exists, update it as $(\cdot,\cdot,\cdot,\csk,\msg)$. If not, add a new entry as $(\cdot, P_i,\cdot,\csk,\msg)$.  
            \item Sample a random $\idx$ and update the entry in $\DDD$ as $(\idx, P_i,\cdot,\csk,\msg)$.
            \item Send $(\mpub, \idx)$ to $\adv$. 
        \end{itemize}

        \item {\bf \retproto}: Upon receiving $(\mret, \csk, \tkn)$ from $P_i$: 
        \begin{itemize}
             \item If $\tkn \notin \LLL_\tknlst$ or $\tkn \in \LLL_\rdmd$ ignore, if  else add $(P_i,\tkn)$ to $\LLL_\rdmd$. 
            \item  If $(\idx,\cdot,\cdot,\csk,\msg)$ exists in $\DDD$, compute $\msg \gets \checkexpiry(\idx,\cdot,\cdot,\csk,\msg)$ and  send $(\mret, \idx)$ to $\adv$. If $(\mret, \ok)$ received from $\adv$ send $(\mret, \csk, \msg)$ to $P_i$ else send  $(\mret, \csk, \bot)$ to $P_i$.
 
        \end{itemize}
       
        \item {\bf Deanonymize Faulter}: Upon receiving $(\faulter, \tkn)$ from some entity, if there exists $>1$ entry of $(P_i, \tkn)$ in $\LLL_\rdmd$, output $(\faulter, P_i)$. 

        \item {\bf Audit Check}: Upon receiving $(\audit,\csk, \msg)$ from $P_i$ , run $\valid(\csk, \msg)$ and return the output.  
    
    \end{itemize}

    \end{tcolorbox}
    \caption{The $\foos$ ideal functionality}
    \label{fig:foos}
\end{figure}
\normalsize

The $\foos$ functionality gives the following interface: 

\begin{itemize}
    \item $\register$: Enables providers to register with the system. This information is leaked to the adversary since it is public. 

    \item $\genid$: A provider with knowledge of some call details - $\callstr$ can request the functionality for a $\csk$. The functionality generates a random $\csk$ and enters an entry associated with this $\csk$ in a database $\DDD$. Notice that the adversary is only notified of a $\genid$ request. It does not learn the corresponding $\callstr$, $\csk$, or the party that invoked this command. This captures the privacy and anonymity guarantees of out-of-band signaling. 

    \item $\mpub$: A provider can submit a message along with a $\csk$. The functionality updates $\DDD$ with the message at the entry corresponding to $\csk$. At this point, the functionality randomly generates an index denoted $\idx$ and sends only this $\idx$ to the adversary. Again, this captures that the adversary does not know the contents of the message $\msg$, nor does it know for what call details this message was submitted, and finally gets no information about the adversary. 

    \item $\mret$: A provider can request to retrieve a message by specifying the $\csk$. The functionality retrieves the corresponding message in $\DDD$ and sends it to the provider. At this point, the server is informed of the $\idx$ but not given information about the identity. 
\end{itemize}

\subsection{The \sysname protocol}
Our protocol requires the following ingredients: 
\begin{enumerate}
    \item Threshold Group Signatures instantiated using the scheme proposed by Bootle et al.~\cite{camenisch2020short} 
    \item An OPRF scheme instantiated using the 2HashDH scheme of Jarecki et al.~\cite{jarecki2017toppss} 
    \item A symmetric encryption scheme 
\end{enumerate}

\sysname consists of  four protocols: \emph{\setupproto}, \emph{\cidgenproto}, \emph{\pubproto}, and \emph{\retproto} as which we describe in details in Figure~\ref{fig:protocolspecs}. Note that we describe the full protocol with distributed $\mev$ and $\mms$.

\begin{figure}
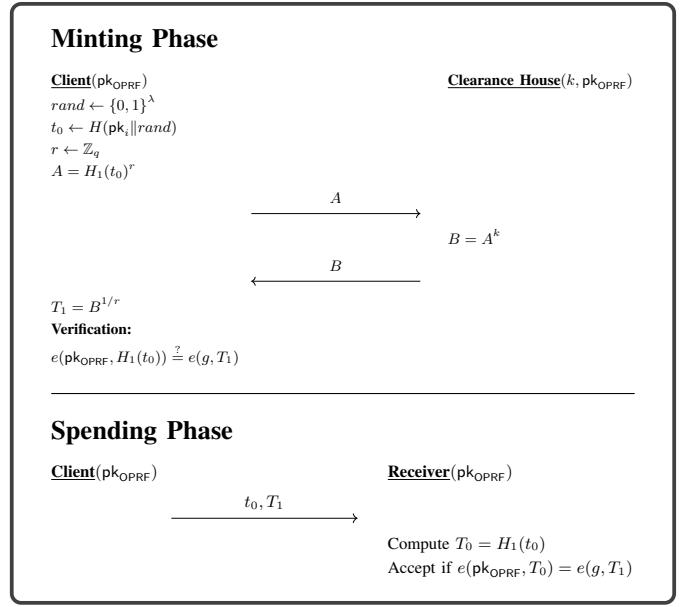

    \centering
    \begin{tcolorbox}[standard jigsaw, opacityback=0]

{\normalsize \textbf{Minting Phase}}

    \vspace{0.5em}
    \begin{adjustbox}{max width=\textwidth, max totalheight=\textheight, center}
    \pseudocode{\textbf{\underline{Client}}(\pk_\oprf) \>\> \textbf{\underline{Clearance House}}(k, \pk_\oprf)\\
        rand \gets \{0,1\}^\lambda\\
        t_0 \gets H(\pk_i\|rand) \\
        r \gets \ZZ_q\\ 
        A = H_1(t_0)^r\\
        \> \sendmessageright*{A} \> \\
        \>\> B = A^k \\
        \> \sendmessageleft*{B}\> \\
        T_1 = B^{1/r} \\
        \textbf{Verification:} \\
        e(\pk_\oprf, H_1(t_0)) \stackrel{?}{=} e(g, T_1)
    }
    \end{adjustbox}

    \vspace{1em}
    \hrule
    \vspace{1em}

{\normalsize \textbf{Spending Phase}}

    \vspace{0.5em}
    \begin{adjustbox}{max width=\textwidth, max totalheight=\textheight, center}
    \pseudocode{\textbf{\underline{Client}}(\pk_\oprf) \>\> \textbf{\underline{Receiver}}(\pk_\oprf)\\
        \> \sendmessageright*{t_0, T_1} \> \\
        \>\> \text{Compute } T_0 = H_1(t_0) \\
        \>\> \text{Accept if } e(\pk_\oprf, T_0) = e(g, T_1)
    }
    \end{adjustbox}

    \end{tcolorbox}
    \caption{Billing Tokens: Minting and Spending Phases }
    \label{fig:billing}
\end{figure}

\ifoobfullversion
\begin{figure*}
\else
\begin{figure}
\fi
\begin{tcolorbox}[standard jigsaw,opacityback=0]
\ifoobfullversion\else\footnotesize\fi
    
    \myparagraph{\setupproto}
    \begin{enumerate} 
        \item \textbf{Group Setup}: Run the $\gkgen$ algorithm of the group signature scheme and announce $\gpk, \info_0$, which are the group manager's public key and initial group information. 
        \item \textbf{Evaluator Setup}: Each Evaluator $\mev_i$ 
        \begin{itemize}
            \item Generates $B$ OPRF secret keys $\{x_{i,j}\}_{j \in [B]}$
            \item Initializes an integer variable $e_i$
            \item Registers with the group manager and receives an $\nodeid_i$. 
        \end{itemize}
        \item \textbf{Message Store Setup}: Each message store $\mms_j$ register with the group manager and receives a $\nodeid_j$. 
        \item \textbf{Provider Setup}: Each $\provider_i$ 
        joins the group by running the interactive protocol $\gjoin$ with the group manager and receives $(\gsk_i, \gpk)$ 
        \item \textbf{Clearinghouse Setup}: Generate $\oprf$ keys - $(\vk_{ch}, k_{ch})$
    \end{enumerate}

    \myparagraph{Minting Coin}: $P_i$ and Clearinghouse run the Mint Phase (Fig~\ref{fig:billing})  and gets back $(t_0^i, T_1^i)_j$ for $j \in \tknlst_i$. 

    All messages from the $\mev$ and $\mms$ are signed under the corresponding signing keys. 
    
    \myparagraph{\cidgenproto}

    Each provider $\provider_i$ with input $\callstr$ 
    
    \begin{enumerate}
        \item Compute $\calldt \gets H(\src\|\dst\|ts)$ 
        \item Compute $i_k \gets \calldt \mod N_{ik}$ 
        \item Pick $r \gets \ZZ_q$ as a mask and compute $a = H_1(\calldt)^r$
        
        \item Compute $\SSS_\mev = \{\mev_1, \ldots, \mev_n\} \gets \findnodes(\calldt, n)$
        \item Let $(t_0^i, T_1^i) \gets\tks_i.\mathsf{get(\calldt)}$
        \item Compute $\sigma \gets \gsign{\gsk_i}{ (a\|i_k\|\SSS_\mev\|(t_0^i, T_1^i))}$
        
        \item Send $(i_k, a, (t_0^i, T_1^i)_j, \SSS_\mev, \sigma)$ to each $\mev_j$ in $\SSS_\mev$.  
\end{enumerate} 

    Each evaluator $\mev_j \in \SSS_\mev$ upon receiving $(i_k, a, (t_0^i, T_1^i), \sigma)$: 
    \begin{enumerate}
        \item Check if $e(\vk_{ch}, t_0^i) = e(g, T_1)$ and abort if not. 
        \item Check $\gverify{\gpk}{(a\|i_k\|\SSS_\mev\|(t_0^i, T_1^i)_j)}{\sigma}$, and abort if it doesnt verify. 
        \item Let $k_{i_k} := \keylist_{j}[i_k]$ Compute $b_j \gets a^{k_{i_k}}$. 
        If $i_k = (k-1) \mod S$, also compute $b_j' = a^{k}$ and return $b_j$ (optionally $b_j'$) to the provider. 
        \item Store $(i_k, a, (t_0^i, T_1^i), \sigma)$. 
\end{enumerate}

    Provider $\provider_i$ upon receiving $b_j$ from $\mev_j$ outputs $\csk \gets \hash(\left( \prod_{j=1}^{\evparam} b_j \right)^{1/r})$.

    \myparagraph{\pubproto}

    Provider $\provider_i$ after computing $\csk$ and with input $\msg$:

        \begin{enumerate}
        \item Compute $\idx \gets \hash(\csk)$
        \item Samples a random string $c_0 \gets \{0,1\}^\lambda$.

        \item Compute $\key_{\mathsf{enc}} \leftarrow  \hash(c_0 \| \csk)$.
    
        \item Compute $c_1 \leftarrow \enc(\key_{\mathsf{enc}}, \msg)$. (one time pad $\key_{\mathsf{enc}} + \msg$)

        \item Compute $ \MMM = \{\mms_1, \ldots, \mms_m\} \gets \findnodes(\csk, m)$ 

\item Let $(t_0^i, T_1^i) \gets\tks_i.\mathsf{get(\calldt)}$
        \item Compute $\sigma = \gsign{\gsk_i}{(\idx\|c_0\|c_1\|\MMM\|t_0^i\| T_1^i}$.

        \item Send  $(\idx, c_0,c_1, (t_0^i, T_1^i), \sigma)$ to the message stores $\{\mms_j\}$ for each $\mms_j\in\MMM$. 
\end{enumerate}
        
         The message store $\mms_j$ upon receiving $(\idx, c_0,c_1, \sigma)$ 
        \begin{enumerate}
            \item Verify $\gverify{\gpk}{(\idx\|c_0\|c_1)}{ \sigma} = 1$ 
            \item Check if $e(\vk_{ch}, t_0^i) = e(g, T_1)$ and abort if not.
            \item Stores $(c_0, c_1, \sigma)$  in a database $\DDD$ at index $\idx$. 
\end{enumerate}

    \myparagraph{\retproto} 
    A provider $\provider_i$ with input $\csk$ does

        \begin{enumerate}
        \item Compute $\idx = \hash(\csk)$. 
        \item Let $(t_0^i, T_1^i) \gets\tks_i.\mathsf{get(\calldt)}$
        \item Compute $\sigma \gets \gsign{\gsk_i}{\idx, (t_0^i\| T_1^i)}$
        \item Compute $ \MMM = \{\mms_1, \ldots, \mms_m\} \gets \findnodes(\csk, m)$
        
        \item Send $(\retreq, (idx,(t_0^i, T_1^i), \sigma))$ to each $\mms_j \in \MMM$. 
\end{enumerate}

       Each storage server $\mms_j$ does: 
        \begin{enumerate}
            \item Abort if $\gverify{\gpk}{\idx_i}{\sigma} == 0$.
                \item Abort if the record indexed at $h$ has expired.
                \item Otherwise Send $(\retres, \idx, (c_0, c_1,\sigma))$ to $\provider_i$ 
        \end{enumerate}
        $P_i$ upon receiving $(\retres, \idx, (c_0, c_1,\sigma))$ does:
        \begin{enumerate}
        \item Verify $\gverify{\gpk}{(\idx\| c_0\| c_1)}{\sigma}$
            \item Compute $k_\mathsf{enc} = \hash(c_0 \| \csk)$ and output $\msg = \dec(k, c_1)$. 
        \end{enumerate}

        \myparagraph{Deanonymize Faulter} The \clearinghouse upon receiving the same tokens $(\tkn, \sigma_1), (\tkn,\sigma_2)$:
        \begin{enumerate}
            \item Sends the corresponding group signatures $\sigma_1,\sigma_2$ to the \SOsLong{s}. 
            \item The \SOsLong{s} run $\gopen(\sigma_1) \rightarrow P_1$ and $\gopen(\sigma_1) \rightarrow P_2$ and punish $P_1, P_2$ (Note that $P_1, P_2$ may be the same provider as well. )
        \end{enumerate}
    
\end{tcolorbox}
\caption{The \sysname protocol}
\label{fig:protocolspecs}
\ifoobfullversion
\end{figure*}
\else
\end{figure}
\fi
\normalsize

\begin{figure}
    \centering
    \footnotesize
    \begin{tcolorbox}[standard jigsaw, opacityback=0]
    The audit logging service maintains lists $\DDD_{cidgen}$, $\DDD_{cidcomp}$, $\DDD_{pub}$, $\DDD_{ret}$
    
        \myparagraph{\cidgenproto Audit Logging} 
        \begin{enumerate}
            \item Upon receiving $(\x, \ik, t_0, t_1, \SSS_\mev, \cpsResPyld, \sigma, \sigma_j)$ check that the signatures verify and store the tuple in $\DDD_{cidgen}$. 
            \item Upon receiving $(\fx_j, \pk_j, \sigma_j)$ check that $\sigma_j$ is valid and $\pk_j$ is the correct corresponding key for $\mev_j$. And store the tuple in $\DDD_{cidcomp}$
        \end{enumerate}
        \myparagraph{Record Publish Audit Logging}: Upon receiving $(H(\idx\|c_0\|c_1), t_0, t_1, \SSS_\mms, \sigma, \sigma_r)$ check that the signatures verify and store the tuple in $\DDD_{pub}$
        \myparagraph{Record Retrieval Audit Logging}: 
        \begin{enumerate}
            \item Upon receiving $(H(\idx), t_0, t_1, \SSS_\mms, \cpsResPyld, \sigma_i, \sigma_r)$ check that the signatures verify and store the tuple in $\DDD_{ret}$
        \end{enumerate}
        \myparagraph{Dispute Resolution}:
        \begin{enumerate}
            
        \item Upon receiving an invalid response complaint from some entity, check if the corresponding $\fx_j$ exists in $\DDD_{cidcomp}$ and check that the pairing equation is valid, and the signature verifies. If both are true, then the corresponding $\mev$ acted honestly, else the $\mev$ is corrupted. 
        
        \end{enumerate}
    \end{tcolorbox}
    \caption{\sysname Audit Logging}
    \label{fig:audit}
    \normalsize
\end{figure}

\subsection{Proof of Security}

In this section, we prove the security of \sysname in the UC model. We will consider the following corruption model: 
A subset of \SOsLong{s}, Message Stores, and Evaluators are corrupt and can collude with a subset of the providers. 

This is the strongest collusion model. If we prove security when these entities are corrupt, then we automatically also have a proof for the case where a subset of the above entities is corrupt. 

To prove security in the UC model, we are required to show a simulator that interacts with the ideal functionality and interacts with the adversary in the real world, and generates a view that is indistinguishable from the real-world view. We describe this simulator next: 

\noindent\underline{\textbf{Simulator $\SSS$}}:

\myparagraph{Setup} 
\begin{enumerate}
    \item Group Manager setup: Run the $\gkgen$ algorithm of the group signature scheme and announce $\gpk, \info_0$, which are the group manager public key and initial group information. 
    \item Honest Evaluator Setup: On behalf of honest evaluators, generate $B$ OPRF secret keys. and publish the corresponding public keys. 
    \item Honest Clearinghouse Setup: On behalf of each honest Clearinghouse, generate OPRF keys and publish the public keys. 
\end{enumerate}

\myparagraph{Register} 
\begin{enumerate}
    \item (Honest Providers) Upon receiving $(\register,P_i)$ from $\foos$, run interactive $\mathbf{Join}$ protocol with corrupt group managers. 
    \item (Corrupt Providers) Upon receiving $\mathbf{Join}$ from a corrupt provider, send $(\register,P_i)$ on behalf of that party to $\adv$. 
\end{enumerate}

\myparagraph{Get Token}
\begin{enumerate}
    \item (Honest Providers) Upon receiving $\gettkn$ from $\foos$, simulate the mint phase of the Billing Protocol with \clearinghouse, sample a random group element and send it to $\foos$. 
    \item (Malicious Providers) Upon receiving the first message $(A)$ from malicious $P_i$, send $\gettkn$ to $\foos$ and send back $B = A^k$ to the malicious $P_i$. 
 \end{enumerate}

\myparagraph{Random Oracle Queries} All random oracle queries are simulated with lazy sampling: for a query $x$, if $H(x)$ is defined, then respond with $H(x)$, else randomly sample $y$, set $H(x) = y$, and respond with $y$. 

\myparagraph{\cidgenproto} We need to simulate the case of an honest provider interacting with a malicious $\mev$, and a malicious provider interacting with an honest $\mev$. 
\begin{enumerate}
    \item (Honest Provider): Upon receiving $\genid$ from $\foos$, sample a random $r \gets \ZZ_q$, and send $a = H_1(0)^r$ to $\adv$ (on behalf of malicious $\mev$ if any) along with a $\tkn = (t_0, T_1)$ that was generated earlier for $P_i$. Upon receiving $b$, compute $\csk_j = b^{1/r}$. 

    \item (Malicious Provider): Now upon receiving $a$ and $\tkn$ from the adversary: 
    \begin{enumerate}
    
        \item Check that the token is valid, if not, ignore
        \item Compute $b = a^k$ and send back $b$ to $\adv$ 
    \end{enumerate}
    
\end{enumerate}

\myparagraph{\pubproto} 
\begin{enumerate}
    \item (Honest Provider) Upon receiving $(\mpub, \idx)$ from $\foos$: 
    \begin{enumerate}
        \item Sample random key and randomly sample $c_0,c_1$  
        \item Send $(\idx, c_0, c_1, \sigma)$ to $\adv$ (on behalf of $\mms$).
    \end{enumerate}

    \item (Corrupt Provider) Upon receiving $(\idx, c_0,c_1, \sigma)$ on behalf of an honest $\mms_j$ 
    \begin{enumerate}
        \item From list of random oracle queries by $\adv$ find the entry with $((c_0 \| \csk), y)$. If such an entry does not exist, output $\mathsf{ROFail}$ and abort. 
        \item Compute $\msg = y - c_1$ 
        \item Send $(\mpub, \csk, \msg)$ to $\foos$. 
    \end{enumerate}
\end{enumerate}

\myparagraph{\retproto} 
\begin{enumerate}
    \item (Honest Provider) Upon receiving $(\mret, \idx)$ from $\foos$ send $(\retreq, \idx, \sigma)$ to $\adv$ on behalf of the message stores. 
    \item (Malicious Provider, Honest $\mms$) Upon receiving $(\retreq, \idx, \sigma)$ from $\adv$, first retrieve the corresponding  $\csk$ from list of random oracle queries, and send $(\mret,\csk)$ to $\foos$. Upon receiving $m$ from $\foos$ do the following: 
    \begin{enumerate}
        \item Sample random string $c_0$
        \item Compute $k_{\mathsf{enc}} = \hash(c_0 \| \csk)$
        \item Compute $c_1 \gets k_{\mathsf{enc}} + m$
        \item Send $(c_0, c_1)$ back to $\adv$
    \end{enumerate}
    \item Now there is also the case that a malicious provider interacts with a malicious $\mms$, while we cannot simulate this interaction, we still need to ensure that the message decrypted by the adversary matches. We achieve this by programming the random oracle appropriately.
    \begin{enumerate}
        \item  Upon receiving a random oracle query $(c_0 \|\csk)$, the simulator first retrieves $(c_0,c_1)$ that was sent to the $\mms$ as part of Record Publish. 
        \item Send $(\mret, \csk)$ to the $\foos$ ideal functionality and receive back $(\mret, \csk, m)$. 
        \item Now the simulator updates the output of the random oracle as follows: $\hash(c_0\|\csk) = y = c_1 - m$. 
    \end{enumerate}
 \end{enumerate}
    \myparagraph{Handling Tokens}
    For any of the interactions, if a $\tkn$ is received
    
    \begin{enumerate}
        \item If this $\tkn$ was generated on behalf of an honest party abort with $\mathsf{TokenFail}_1$.
        \item If this $\tkn$ was not generated by the simulator on behalf of \clearinghouse abort with $\mathsf{TokenFail_2}$
    \end{enumerate} 

    \myparagraph{Handling Group Signatures} For any of the interactions if a group signature $\sigma$ was received: First compute $\gopen(\sigma)$ and output $P_i$. If $P_i$ corresponds to an honest party abort with $\mathsf{GroupSigFail}$. 

    \myparagraph{Deanonymize Faulter} Upon receiving two group signatures and the same token from some honest party, send $(\faulter, \tkn)$ to the $\foos$ ideal functionality, and output whatever the functionality outputs. 

    \myparagraph{Simulating Audit Logging} The simulation follows exactly as the protocol, except that the simulation aborts if a valid signature is received from an entity on behalf of an honest party.

Now that we have described the simulator, we show via a sequence of hybrids that the real world and the simulated world are indistinguishable.

\noindent\underline{\textbf{Proof By Hybrids}}: 

$\hyb_0$: This is the real-world protocol. 

\vspace{4pt} 

$\hyb_1$: This hybrid is the same as the previous hybrid except that the simulator may abort with the error $\mathsf{TokenFail}_1$. Since we assume an unforgeable OPRF, the probability that this event occurs is negligible, and therefore, the two worlds are indistinguishable. 

\vspace{4pt}

$\hyb_2$: This hybrid is the same as the previous hybrid, except that the simulator aborts with $\mathsf{TokenFail}_2$. Again, since we assume an unforgeable OPRF, the probability that this event occurs is negligible, and therefore the two worlds are indistinguishable. 

\vspace{4pt}

$\hyb_3$: This hybrid is the same as the previous hybrid, except that the simulator aborts with the message $\mathsf{GroupSigFail}$. By the non-frameability property of the underlying group signature scheme, this event occurs with negligible probability, and these two hybrids are indistinguishable. 

\vspace{4pt}

$\hyb_4$: This hybrid is the same as the previous hybrid except that the simulator output $\mathsf{ROFail}$. Since we use the random oracle for all hash function queries and require that the adversary uses the random oracle as well, this event occurs with negligible probability. 

\vspace{4pt}

$\hyb_5$: This hybrid is the same as the previous hybrid except that we replace the OPRF queries to the evaluators with 0. By the obliviousness property of the underlying OPRF scheme, these two hybrids are indistinguishable. 

\vspace{4pt}

$\hyb_6$: This hybris is the same as the previous hybrid except that ciphertexts $(c_0, c_1)$ are replaced by random strings. Since we use the one-time pad for $\enc$ and the distribution of the ciphertexts is not affected, these two hybrids are indistinguishable.

\vspace{4pt}

$\hyb_7$: This hybrid is the same as the previous hybrid except that the simulator aborts in the audit logging phase upon receiving valid signatures on behalf of honest parties. Since we use an unforgeable signature, the abort events occur with negligible probability, and the two hybrids are indistinguishable. 

Since this hybrid is identical to the simulated world, we have shown that the real-world and the ideal world are indistinguishable, and that concludes the proof of security of our scheme. 

  \subsection{\sysname Inter-Work Function (\iwf)}\label{sec:iwf}

\begin{figure*}[t!]
  \begin{minipage}[t]{0.99\textwidth}
    \centering
    \captionsetup[subfigure]{justification=centering}
    {\includegraphics[width=0.75\linewidth]{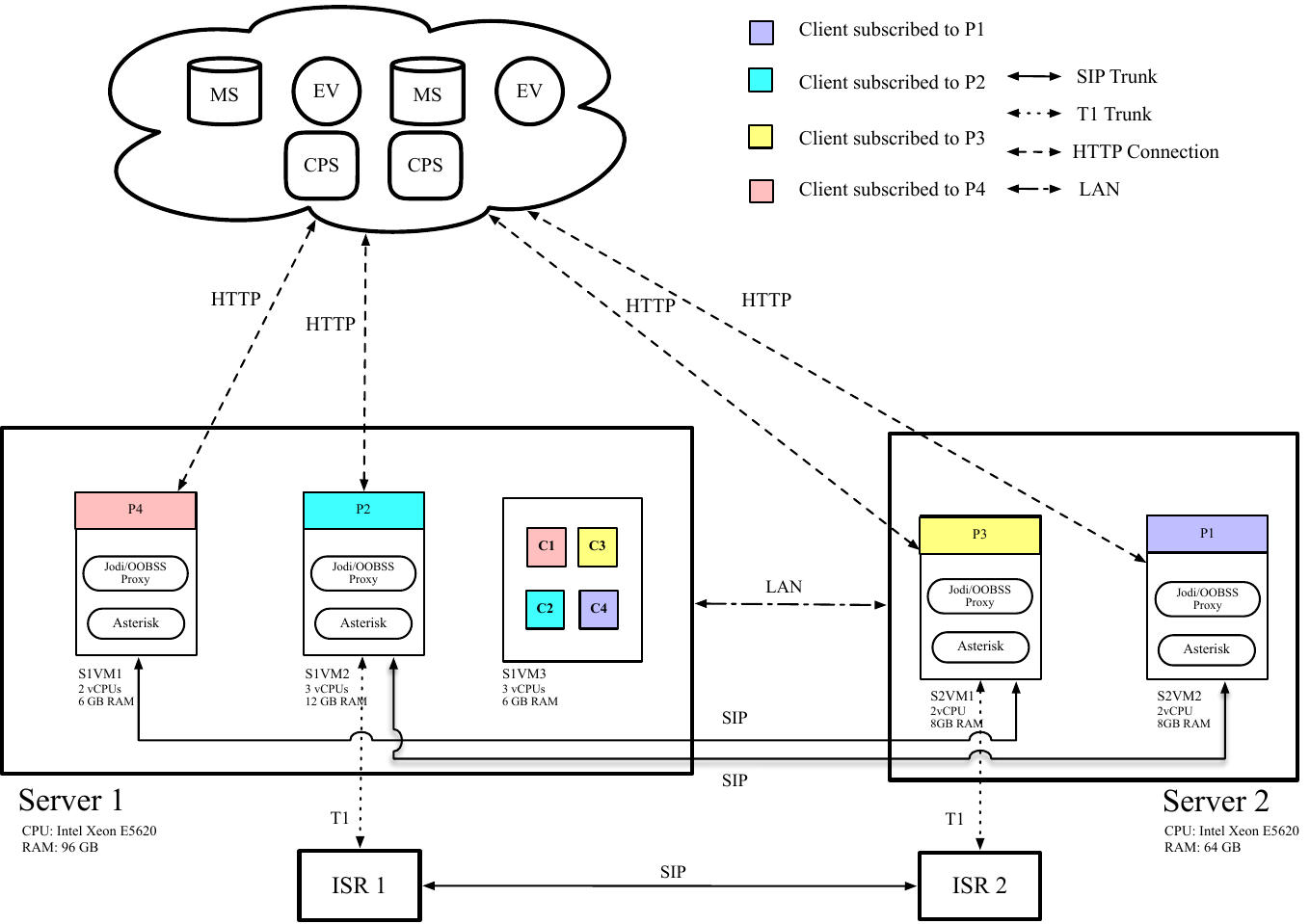}}
    \hfill
    
    \caption{Testbed Architecture}
    \label{fig:testbed-arch}
  \end{minipage}
\end{figure*}

\iwf consists of a lightweight proxy server implemented in \texttt{FastAPI}, packaged as a Docker container, and a single-header C library that exposes two core functions: \texttt{publish} and \texttt{retrieve}. Providers integrate the library directly into their gateways to enable passport publication and retrieval.

We integrated \iwf into \texttt{Asterisk}, a widely used open-source PBX platform that already supports \stirsha attestation and verification. To support realistic TDM functionality, we extended \texttt{Asterisk} using \texttt{Sangoma A102 T1/E1 PCIe} cards and two \texttt{Cisco ISR 4331} routers to facilitate TDM/ISDN trunk traffic. 

Fig.~\ref{fig:testbed-arch} shows our testbed architecture with four provider nodes deployed across two physical servers. Each server runs virtual machines hosting providers running our extended \texttt{Asterisk}.

As shown, $\provider_1$ connects to $\provider_2$ via a SIP trunk. $\provider_2$ connects to $\provider_3$ through two ISRs, forming both SIP and TDM trunks, and $\provider_3$ connects to $\provider_4$ via another SIP trunk. $\provider_2$ connects to \texttt{ISR 1} using a \texttt{T1} trunk powered by \texttt{Sangoma A102 T1/E1 PCIe cards}, and $\provider_3$ connects to \texttt{ISR 2} in the same way.

When $\provider_1$ initiates a call to $\provider_4$, the call flows from $\provider_1$ to $\provider_2$ over SIP, converts to TDM, passes through $ISR_1$, converts back to SIP, passes through $ISR_2$, converts again to TDM, then reaches $\provider_3$ and finally $\provider_4$ via SIP. The provider converting from SIP to TDM ($\provider_2$) uploads the \passport using \iwf's \texttt{publish} capability, and the provider converting back to SIP ($\provider_3$) downloads it using the \texttt{retrieve} capability.

We simulate clients on server 1 using the \texttt{PJSIP2} library. Caller $C_2$ connects to $\provider_2$, and the callee $C_3$ connects to $\provider_3$.

\end{document}